\documentclass[11 pt,oneside,onecolumn,a4paper]{article}
\usepackage{epsfig,graphicx}
\usepackage{cite}
\usepackage{indentfirst}
\usepackage{amsfonts,amssymb,latexsym,amsmath,enumerate,amsthm}
\usepackage{indentfirst,cite}
\usepackage{bm}
\usepackage{amsmath}
\usepackage{amsmath}

\def\det{{\rm det}}
\def\sgn{{\rm sgn}}

mm \leftmargin 5pt \rightmargin 5pt \hoffset= -15mm

\title{Vortex particles in axially symmetric fields \\ and applications of the quantum Busch theorem} 

\author{Dmitry Karlovets}

\date{\small{Faculty of Physics, Tomsk State University, \\ Lenina Ave.\,36, 634050 Tomsk, Russia}}

\begin{document}

\maketitle

\begin{abstract}
The possibilities to accelerate vortex electrons with orbital angular momentum (OAM) to relativistic energies and to produce vortex ions, protons, and other charged particles
crucially depend on whether the OAM is conserved during the acceleration and on how phase space of the wave packet evolves.
We show that both the OAM and a mean emittance of the packet, the latter obeying the Schr\"odinger uncertainty relation, are conserved in axially symmetric fields 
of electric and magnetic lenses, typical for accelerators and electron microscopes, as well as in Penning traps, while a linear approximation of weakly inhomogeneous fields 
works much better for single packets than for classical beams. We analyze quantum dynamics of the packet's rms radius $\langle\rho^2\rangle$,
relate this dynamics to a generalized form of the van Cittert–Zernike theorem, applicable at arbitrary distances from a source and for non-Gaussian packets,
and adapt the Courant-Snyder formalism to describe the evolution of the wave packet's phase space. The vortex particles can therefore be accelerated, focused, steered, trapped, and even stored in azimuthally symmetric fields and traps, somewhat analogously to the classical angular-momentum-dominated beams. Moreover, we give a quantum version of the Busch theorem, which states how one can produce vortex electrons with a magnetized cathode during either field- or photoemission, as well as vortex ions and protons by using a magnetized stripping foil employed to change a charge state of ions. Spatial coherence of the packets plays a crucial role in these applications and we provide the necessary estimates for particles of different masses. 

\end{abstract}


\section{Introduction}

The vortex electrons carrying orbital angular momentum (OAM) $\ell\hbar$ with respect to their mean momentum have potential applications in atomic physics, particle physics, hadronic and spin physics \cite{Bliokh, Lloyd, Ivanov11, Ivanov12, PRA12, Bliokh12, McMorran, Beche2014, FA1, FA2, Serbo15, JHEP, Ivanov16, PRA18, Moments, PRA19, PRD, SilenkoFields, SilenkoQ, Silenko2020, Sherwin1, Sherwin2, Ivanov20201, Ivanov20202, Ivanov20203}. In many of these applications, relativistic energies are needed, which are much higher than the current upper limit of $\varepsilon_c \approx 300$ keV achieved with electron microscopes \cite{Uchida, Verbeeck, McMorran2}. Although quantum dynamics of the vortex electrons in electromagnetic fields has been extensively studied in recent years \cite{PRA12, McMorran, Bliokh12, Beche2014,  FA1, FA2, SilenkoFields, SilenkoQ, Silenko2020}, there is still no clear understanding of whether the OAM is conserved in electric and magnetic lenses, employed in accelerators and electron microscopes for focusing and phase-space manipulations. Here we address this question by pointing out that the OAM of a vortex particle is conserved in a wide class of azimuthally symmetric electric and magnetic fields, while a transverse emittance of the packet is also conserved on average in weakly inhomogeneous -- that is, linear -- fields, analogously to that of a classical beam \cite{Reiser}. This suggests means for accelerating the vortex electrons to relativistic energies.

Generation of vortex states of heavier charged particles, such as light and heavy ions, protons, antiprotons, and so on, would stimulate studies of new wave phenomena in atomic and 
molecular physics, heavy ion physics, high-energy physics. The possibilties to put a twist on such composite particles heavily depend on their coherence properties and on temperature of their sources.
Remarkably, here is where physics of classical beams in accelerators can suggest methods for the generation of wave packets with definite quantum numbers.
The so-called angular-momentum-dominated beams of electrons, ions, and protons represent a classical counterpart of the vortex states and they are routinely produced via cathodes and stripping foils immersed into a solenoid field \cite{Burov2000, Burov2002, Kim2003, Sun2004, Groening2014, Hwang2014, Groening2017, Groening2018}. This technique invokes a classical Busch theorem \cite{Busch} and it allows one to generate beams with the very large unquantized angular momenta, while also redistributing the phase space and switching between a flat and a round beam, 
analogously to the mode conversion in optics \cite{Scht, Floettmann2020}. As we have recently noted \cite{PRA2020}, this method can be adapted to generate quantum vortex states.

Analogies between optics and quantum mechanics due to similarities of the paraxial wave equation and the Schr\"odinger equation are well known. 
The ideas and methods of \textit{particle optics} in accelerators can also be borrowed from classical into the quantum domain \cite{Manko98, Manko99, Dodonov2000} due to similarities between the Bohmian quantum mechanics \cite{Bohm, Bohm2} and the classical statistical physics. In this paper, we demonstrate that \textit{spatial coherence} of the wave packets and its quantum dynamics play a key role for possibilities to accelerate vortex electrons and to generate twisted states of heavier charged particles. The unnormalized Bessel beams are inapplicable for these purposes and we employ 
the Laguerre-Gaussian packets \cite{PRA18, PRA19}, adapt the Courant-Snyder formalism of the accelerator physics \cite{Reiser}, 
and apply a concept of \textit{emittance} to a quantum wave packet, which is limited from below by \textit{the Schr\"odinger uncertainty relation} \cite{Sch}. 
The evolution of the particle's rms radius is shown to be governed by a generalized form of the well-known van Cittert–Zernike theorem \cite{Mandel, Cho, Cho2013, Lat, Ehberger},
while the mean emittance is conserved together with the OAM in a wide range of azimuthally symmetric electric and magnetic lenses and also in Penning traps.

We demonstrate how the classical Busch theorem can be adapted into the quantum realm and how it defines the OAM of electrons, ions, protons, etc. born inside a solenoid during field emission, photoemission, 
or as a result of changing the charge state by a stripping medium. The resultant OAM is due to quantized flux of the magnetic field through an area of the wave packet itself at a point of generation, 
which is somewhat similar to the Aharonov-Bohm effect \cite{AB} or to interaction with an effective magnetic monopole \cite{Beche2014}. 

In Sec.\ref{Cl.}, we give a benchmark example of a classical particle in a magnetic field, which is a basic model for the angular-momentum-dominated beams. 
In Sec.\ref{Free}, we study quantum dynamics of a freely propagating quantum packet of an arbitrary shape and show how such a classical concept as emittance can be applied to a wave packet. We develope the Courant-Snyder formalism in Sec.\ref{SecCS} to describe the evolution of the packet's rms radius $\langle\rho^2\rangle$, show its connection to a Gouy phase, and to a generalized form of the van Cittert–Zernike theorem, see Sec.\ref{SecCZ}. These definitions are examplified with the so-called \textit{standard and elegant} Hermite-Gaussian and Laguerre-Gaussian packets. In Sec.\ref{SecSol} we study quantum dynamics of a vortex electron first in a magnetic field and then, in Sec.\ref{SecEnSol} and Sec.\ref{Symm}, in more realistic weakly inhomogeneous fields of solenoids and electric lenses of a finite length.
The applications of the quantum Busch theorem are given in Sec.\ref{BuschEl} for electrons and in Sec.\ref{BuschI} for ions and other charged particles.
We conclude in Sec.\ref{Concl}. When accelerated to ultrarelativistic energies, the transverse dynamics of a packet stays non-relativistic in the overwhelming majority of practical cases, 
which is implied in our analysis. We neglect the spin for the same reason and also because its contribution is marginal for large OAM, $\ell \gg 1$.
The sign of the elementary charge $e$ is arbitrary, the magnetic field stength is positive, $H > 0$, and we also put $\hbar = c = 1$.

\section{Benchmark example: classical particle in magnetic field}\label{Cl.}

A canonical (generalized) momentum of a point charged particle in a field with a potential ${\bm A}$ is
\begin{eqnarray}
& \displaystyle {\bm p}^{\text{can}} = {\bm p} + e {\bm A}.
\label{pdef}
\end{eqnarray}
For a homogeneous and constant magnetic field we have
\begin{eqnarray}
& \displaystyle {\bm A} = \frac{1}{2}\, [{\bm H} \times {\bm r}] = \frac{H}{2}\,\{-y, x, 0\} = \frac{H\rho}{2}\,{\bm e}_{\phi},\cr
& \displaystyle {\bm H} = \{0,0,H\},\, H > 0.
\label{Adef}
\end{eqnarray}
The general solutions to the classical equations of motion are \cite{LL2},
\begin{eqnarray}
& \displaystyle {\bm r} = {\bm r}_0 + \{\rho_c \sin(\omega_ct + \alpha), \rho_c \cos(\omega_ct + \alpha), u_z t\},\cr
& \displaystyle {\bm p} = {\bm u} m = \{p_{\perp} \cos(\omega_ct + \alpha), -p_{\perp} \sin(\omega_ct + \alpha), p_z\},
\label{rp}
\end{eqnarray}
where ${\bm r}_0$ and $\alpha$ are defined by initial conditions and
\begin{eqnarray}
& \displaystyle \rho_c = \frac{p_{\perp}}{eH} = \frac{u_{\perp}}{\omega_c},\quad \omega_c = \frac{eH}{m}.
\label{rc}
\end{eqnarray}
The absolute values $|\rho_c|$ and $|\omega_c|$ are called a cyclotron radius and a cyclotron frequency, respectively.

In such a problem, there exists a so-called adiabatic invariant \cite{LL2},
\begin{eqnarray}
& \displaystyle I = \oint\frac{d{\bm \rho}}{2\pi}\,{\bm p}^{\text{can}}_{\perp} = \oint \frac{d{\bm \rho}}{2\pi}\, {\bm p}_{\perp} + e \oint \frac{d{\bm \rho}}{2\pi}\,{\bm A},
\label{I}
\end{eqnarray}
where the integration is taken over the particle's path. The invariant is equal to
\begin{eqnarray}
& \displaystyle I = \frac{{\bm p}_{\perp}^2}{2eH} = \omega_L m \rho_c^2 = \frac{\varepsilon_{\perp}}{\omega_c} = \text{inv},
\label{I2}
\end{eqnarray}
where
\begin{eqnarray}
& \displaystyle
\varepsilon_{\perp} = \frac{{\bm p}_{\perp}^2}{2m}\quad \text{and}\quad \omega_L = \frac{\omega_c}{2} = \frac{eH}{2m}.
\label{omegac}
\end{eqnarray}
The absolute value $|\omega_L|$ is called a Larmor frequency.

In fact, each of the two terms in (\ref{I}) conserve separately, as
\begin{eqnarray}
& \displaystyle \oint\frac{d{\bm \rho}}{2\pi}\,{\bm p}_{\perp} = 2I,\  e \oint \frac{d{\bm \rho}}{2\pi}\,{\bm A} = -I.
\label{Isep}
\end{eqnarray}
The invariant $I$ can also be written in terms of a conserving flux $\Phi$ of the magnetic field through a circle of the radius $|\rho_c|$,
\begin{eqnarray}
& \displaystyle I = - e \oint \frac{d{\bm \rho}}{2\pi}\,{\bm A} = \frac{e}{2\pi}\, \Phi,\ \Phi = \int d{\bm s}\, {\bm H} > 0,\, d{\bm s} \uparrow\uparrow {\bm H}.
\label{Iflux}
\end{eqnarray}
The flux is also connected to the $z$-component of the kinetic angular momentum (AM),
\begin{eqnarray}
& \displaystyle L_z = [{\bm r} \times {\bm p}]_z = [{\bm r}_0 \times {\bm p}]_z - 2 I,
\label{Lz}
\end{eqnarray}
where we have used (\ref{rp}). It is generally not $L_z$ itself, but its \textit{intrinsic} -- that is, independent from the origin -- value,
\begin{eqnarray}
& \displaystyle L_z^{\text{int}} = [({\bm r} - {\bm r}_0) \times {\bm p}]_z = - 2 I = \cr
& \displaystyle = 2 e \oint \frac{d{\bm \rho}}{2\pi}\,{\bm A} = -\frac{e}{\pi} \Phi = -\frac{\varepsilon_{\perp}}{\omega_L},
\label{Lzint}
\end{eqnarray}
which is conserved. The intrinsic value of the canonical AM is also conserved,
\begin{eqnarray}
& \displaystyle L_z^{\text{can(int)}} = [({\bm r} - {\bm r}_0) \times {\bm p}^{\text{can}}]_z = \frac{L_z^{\text{int}}}{2} = - I,
\label{Lzcanint}
\end{eqnarray}
and is equal -- up to the sign -- to the adiabatic invariant.

Let us give quantitative estimates of the canonical AM as a function of the field strength and the particle velocity.
To this end, we employ the following representation:
\begin{eqnarray}
& \displaystyle L_z^{\text{can(int)}} = -\frac{\hbar}{2}\, \sgn(e)\, {\bm u}_{\perp}^2 \frac{H_c}{H},
\label{Lzint1}
\end{eqnarray}
where we have introduced a critical magnetic field,
\begin{eqnarray}
& \displaystyle H_c = \frac{m^2}{|e|\hbar},
\label{Hc}
\end{eqnarray}
which is $4.4\cdot 10^{13}\, \text{G} = 4.4\cdot 10^{9}\, \text{T}$ for electron. Clearly, $|L_z^{\text{can(int)}}|/\hbar$ is \textit{not an integer}. 
For the typical magnetic fields of solenoids and magnets $H \sim 0.1 - 1$ T and $u_{\perp} \sim 10^{-2}$, we have
\begin{eqnarray}
& \displaystyle |L_z^{\text{can(int)}}| \sim \hbar\, (10^5 - 10^6),
\label{Lzintabs}
\end{eqnarray}
whereas the corresponding cyclotron radius is
\begin{eqnarray}
& \displaystyle |\rho_c| \sim 1\, \mu \text{m} - 10\, \mu \text{m}.
\label{rcabs}
\end{eqnarray}
The larger radii correspond to the weaker field and therefore to even larger values of the angular momentum.
Thus $L_z^{\text{can(int)}}$ represents an adiabatic invariant in slowly varying or weakly inhomogeneous fields, such as those of solenoids. 
In contrast to the quantum picture below, both the angular momenta vanish when the motion is rectilinear, $\rho_c \rightarrow 0$.

\section{Quantum dynamics of a free wave packet}\label{Free}
\subsection{$1+1$ dimensional space-time}\label{11}
\subsubsection{Emittance and the Schr\"odinger uncertainty relation}

Let us consider a non-relativistic and not necessarily Gaussian packet of a freely propagating massive particle in $1+1$ dimensional space-time. 
Its mean coordinate follows the classical trajectory,
\begin{eqnarray}
& \displaystyle \langle x\rangle = \langle u\rangle\,t,\, \langle u\rangle = \frac{\langle \hat{p}\rangle}{m},
\label{xmean}
\end{eqnarray}
where $\hat{p} = -i\partial_x$ in the coordinate representation, and the Hamiltonian is
\begin{eqnarray}
& \displaystyle \hat{H}_{\text{fr}} = \frac{\hat{p}^2}{2m}.
\label{H1D}
\end{eqnarray}
The latter does not commute with $x^2$ and the Heisenberg's equation of motion is
\begin{eqnarray}
& \displaystyle \frac{\partial^2\langle x^2\rangle(t)}{\partial t^2} = -\left\langle\left[\left[\hat{x}^2,\hat{H}_{\text{fr}}\right],\hat{H}_{\text{fr}}\right]\right\rangle 
= 2 \langle u^2\rangle = 2\frac{\langle\hat{p}^2\rangle}{m^2}.
\label{Heis1D}
\end{eqnarray}
The general solution to this equation reads (cf. \cite{Mess, Manko98})
\begin{eqnarray}
& \displaystyle \langle x^2\rangle(t) = \langle x^2\rangle(0) + \frac{\partial\langle x^2\rangle(0)}{\partial t}\, t + \langle u^2\rangle\, t^2,
\label{xfree}
\end{eqnarray}
and describes spreading of a packet of an arbitrary shape with time and distance.

Let the packet be in a quantum state described with a wave function $\psi$ or with a quantum phase-space distribution (Wigner) function.
We denote $\langle...\rangle$ a quantum mechanical averaging and also
\begin{eqnarray}
& \displaystyle q_i = \{x,u\},\, i = 1,2.
\label{q}
\end{eqnarray}
Let us define a matrix of central moments as follows:
\begin{eqnarray}
Q_{ij} = \langle q_i q_j\rangle - \langle q_i\rangle \langle q_j\rangle.
\label{Q}
\end{eqnarray}
The square root of its determinant, 
\begin{eqnarray}
& \displaystyle \epsilon_x = \sqrt{\det Q_{ij}} = \sqrt{\left(\langle x^2\rangle - \langle x\rangle^2\right) \left(\langle u^2\rangle - \langle u\rangle^2\right) - \left(\langle xu\rangle - \langle x\rangle \langle u\rangle\right)^2}
\label{em1D}
\end{eqnarray}
we call \textit{a normalized emittance} of the wave packet. This general definition is \textit{Lorentz invariant} and it also holds for the particle in external electromagnetic fields. 
For a free particle, we take Eq.(\ref{xfree}) into account and have
\begin{eqnarray}
& \displaystyle \epsilon_x^{\text{free}} = \sqrt{\langle x^2\rangle(0) \left(\langle u^2\rangle - \langle u\rangle^2\right) - \left(\langle xu\rangle(0)\right)^2},
\label{em1Do}
\end{eqnarray}
i.e., the emittance is conserved in time. 

The usefulness of the emittance (\ref{em1D}) is that it explicitly measures non-classicality of the packet via \textit{the Schr\"odinger uncertainty relation}\footnote{Also called the Schr\"odinger-Robertson relation although the work of Robertson \cite{Rob} does not contain the anticommutator in the r.h.s.} \cite{Sch}. Indeed, let us define the following \textit{quality factor} of a wave packet as a measure of non-classicality:
\begin{eqnarray}
& \displaystyle M = \frac{2\epsilon_x}{\lambda_c} \geq 1,
\label{Mwp}
\end{eqnarray}
where
\begin{eqnarray}
& \displaystyle
\lambda_c = \frac{1}{m} \equiv \frac{\hbar}{mc}
\label{lc}
\end{eqnarray}
is a Compton wavelength of the particle with a mass $m$, which is $3.9\cdot 10^{-11}\, \text{cm}$ for electron. We will also need an electron's \textit{Compton time},
\begin{eqnarray}
& \displaystyle
t_c = \lambda_c/c \approx 1.3\times 10^{-21}\, \text{sec}.
\label{tc}
\end{eqnarray}
which is a typical lifetime of a virtual electron-positron pair. Then we have (cf. \cite{Feist})
\begin{eqnarray}
& \displaystyle \frac{M}{2} =  \frac{\epsilon_x}{\lambda_c} = \sqrt{\left(\langle x^2\rangle - \langle x\rangle^2\right) \left(\langle p^2\rangle - \langle p\rangle^2\right) - \left(\langle xp\rangle - \langle x\rangle \langle p\rangle\right)^2} \geq \frac{1}{2},\cr
& \displaystyle \text{or}\quad \epsilon_x \geq \frac{\lambda_c}{2},
\label{em1D2}
\end{eqnarray}
which is just the Schr\"odinger uncertainty relation \cite{Sch}. 
The quality factor $M$ defines the conserved total number of the mutually orthogonal states in the particle's phase space.
The higher this number is, the larger is the r.h.s of the uncertainty relation and the ``more quantum'' is the packet. 
Clearly, even for the simplest free packet it is the Schr\"odinger uncertainty relation that adequately measures the packet's non-classicality 
and not the Heisenberg relation without the correlation term.

\subsubsection{Gaussian packet and the Gouy phase}

As a simplest example, we take a Gaussian packet with a mean momentum $\langle p\rangle$ and its uncertainty $\sigma_p$,
\begin{eqnarray}
& \displaystyle \psi (p) = N\, \exp \left\{-\frac{(p-\langle p\rangle)^2}{2\sigma_p^2}\right\}.
\label{HGp0}
\end{eqnarray}
In the coordinate space we have 
\begin{eqnarray}
& \displaystyle \psi (x,t) = \int\frac{dp}{2\pi}\, \psi (p)\, \exp\left\{-it\frac{p^2}{2m} + ipx\right\} = \cr
& \displaystyle = N\,\frac{1}{\sqrt{\sigma_x(t)}}\, \exp \left\{-it\frac{\langle p\rangle^2}{2m} + i\langle p\rangle x -\frac{i}{2}\, \arctan\frac{t}{t_d} - \frac{(x - \langle u\rangle t)^2}{2\sigma_x^2(t)} \left(1 - i\frac{t}{t_d}\right)\right\},
\label{HGx0}
\end{eqnarray}
where
\begin{eqnarray}
& \displaystyle \sigma_x(t) = \sigma_p^{-1}\sqrt{1 + \frac{t^2}{t_d^2}},\ t_d = \frac{m}{\sigma_p^2},\, \langle x^2\rangle = \frac{\sigma_x^2(t)}{2},\ \frac{\partial\langle x^2\rangle(0)}{\partial t} = 0.
\label{HGx0not}
\end{eqnarray}
The normalization constant $N$ can be found from the condition
\begin{eqnarray}
& \displaystyle \int dx\, |\psi (x,t)|^2 = \int \frac{dp}{2\pi}\, |\psi(p)|^2 = 1.
\label{norm}
\end{eqnarray}
Eq.(\ref{HGx0}) represents an exact \textit{non-stationary} solution to the Schr\"odinger equation,
\begin{eqnarray}
& \displaystyle \left(i\partial_t - \frac{\hat{p}^2}{2m}\right)\psi (x,t) = 0.
\label{Sch0}
\end{eqnarray}

The Gouy phase in (\ref{HGx0})
\begin{eqnarray}
& \displaystyle
\varphi^{G} = \arctan t/t_d
\label{phiG}
\end{eqnarray}
comes about due to the packet's spreading with time, even at rest with $\langle u\rangle$=0.
Note the following identity:
\begin{eqnarray}
& \displaystyle 1 - i\frac{t}{t_d} = \sqrt{1 + \frac{t^2}{t_d^2}}\, \exp\left\{-i\arctan\frac{t}{t_d}\right\}.
\label{G0}
\end{eqnarray}
Importantly, the diffraction (spreading) time $t_d = m/\sigma_p^2$ and the Gouy phase itself \textit{are Lorentz invariant} because the momentum uncertainty $\sigma_p$ and the time $t$ are defined in the packet's rest frame. An inversion of time $t \to -t$ implies $\langle p\rangle \to - \langle p\rangle$ and so (see also \cite{PRA19})
\begin{eqnarray}
& \displaystyle t \to -t:\ \psi (x,t) \to \psi^* (x,t),
\label{tinv}
\end{eqnarray}
in accord with the Wigner theorem and the general CPT-theorem \cite{BLP}.

\subsubsection{Elegant Hermite-Gaussian packet}

One can ``excite'' the Gaussian states by adding a factor $(p - \langle p\rangle)^n$ with $n = 0,1,2,...$ to Eq.(\ref{HGp0}):
\begin{eqnarray}
& \displaystyle \psi_n (p) = N\,(p - \langle p\rangle)^n\, \exp \left\{-\frac{(p-\langle p\rangle)^2}{2\sigma_p^2}\right\}.
\label{HGpn}
\end{eqnarray}
The corresponding wave function of the coordinate representation reads
\begin{eqnarray}
& \displaystyle \psi_n (x,t) = N\, \frac{1}{(1 + t^2/t_d^2)^{\frac{n+1}{4}}}\, H_n\left((x - \langle u\rangle t) \sqrt{\frac{1 - it/t_d}{2\sigma_x^2(t)}}\right)\cr 
& \displaystyle \times \exp \left\{-it\frac{\langle p\rangle^2}{2m} + i\langle p\rangle x -\frac{i}{2}\, (n+1) \arctan\frac{t}{t_d} - \frac{(x - \langle u\rangle t)^2}{2\sigma_x^2(t)} \left(1 - i\frac{t}{t_d}\right)\right\},
\label{HGxn}
\end{eqnarray}
where $H_n$ are Hermite polynomials and a prefactor of the Gouy phase now depends on $n$. One can calculate the following averages:
\begin{eqnarray}
& \displaystyle \langle x^2\rangle = \sigma_x^2(t)\, (n+1/2) - \sigma_x^2(0)\, \frac{n(n-1)}{n-1/2} + \langle u\rangle^2 t^2,\cr
& \displaystyle \langle p^2\rangle = \langle p\rangle^2 + \sigma_p^2\, (n+1/2),\ \langle \varepsilon \rangle = \frac{\langle p^2 \rangle}{2m} = \frac{ \langle p \rangle^2}{2m} + \frac{\sigma_p^2}{2m}\, (n+1/2),\cr
& \displaystyle \langle x u\rangle = \frac{1}{2} \frac{\partial \langle x^2\rangle}{\partial t} = \frac{1}{m}\, \frac{t}{t_d}\, (n+1/2) + \langle u \rangle^2 t,\ \langle x u\rangle(0) = 0.
\label{HGav}
\end{eqnarray}
Note that the use of momentum represenation is somewhat easier here, and the coordinate operator in momentum space is
\begin{eqnarray}
& \displaystyle
\hat{x} = i\frac{\partial}{\partial p} + \frac{p}{m} t.
\label{xp}
\end{eqnarray}
Although this packet is non-stationary and spreads, its mean energy $\langle \varepsilon \rangle$ is conserved.
 
The emittance of this packet is
\begin{eqnarray}
& \displaystyle \epsilon_x = \frac{\lambda_c}{2}\, \sqrt{\frac{(2 n + 1) (4 n - 1)}{2 n - 1}},
\label{emx}
\end{eqnarray}
whereas the quality factor reads
\begin{eqnarray}
& \displaystyle M = \frac{2\epsilon_x}{\lambda_c} = \sqrt{\frac{(2 n + 1) (4 n - 1)}{2 n - 1}} = \left\{
    \begin{array}{ll}
        1,\ n=0, \\
        2\sqrt{n},\ n \gg 1.
    \end{array}
		\right.
\label{M2x}
\end{eqnarray}

Eq.(\ref{HGxn}) also represents an exact non-stationary solution to the Schr\"odinger equation and in optics it is often called an \textit{elegant} Hermite-Gaussian (HG) packet \cite{Siegman,Siegman97,Lu}.
The set of these functions, however, is not orthogonal, 
\begin{eqnarray}
& \displaystyle \int dx\, \psi_n^* (x,t) \psi_m (x,t) \not\propto \delta_{nm},
\label{HGElOrth}
\end{eqnarray}
but rather \textit{biorthogonal} as the function adjoint to $\psi_n(x,t)$ is not $\psi_n^*(x,t)$ \cite{Siegman,Siegman97}.

\subsubsection{Standard Hermite-Gaussian packet}

There exists a very similar but \textit{not identical} set of functions, which is complete and orthogonal and it can be achieved by the following substitution in Eq.(\ref{HGpn}):
$$
(p - \langle p\rangle)^n \rightarrow \left(p - \langle p\rangle - i\, \text{const}\,\frac{\partial}{\partial p}\right)^n
$$ 
The corresponding coordinate wave function
\begin{eqnarray}
& \displaystyle \psi_n (x,t) = N\, \frac{1}{(1 + t^2/t_d^2)^{1/4}}\, H_n\left(\frac{x - \langle u\rangle t}{\sigma_x(t)}\right)\cr 
& \displaystyle \times \exp \left\{-it\frac{\langle p\rangle^2}{2m} + i\langle p\rangle x -\frac{i}{2}\, (2 n+1) \arctan\frac{t}{t_d} - \frac{(x - \langle u\rangle t)^2}{2\sigma_x^2(t)} \left(1 - i\frac{t}{t_d}\right)\right\}
\label{HGxnSt}
\end{eqnarray}
also satisfies the Schr\"odinger equation exactly and it is called a \textit{standard} Hermite-Gaussian packet. Compared to Eq.(\ref{HGxn}), a prefactor of the Gouy phase has now $2n+1$ instead of $n+1$, which is closely connected with the quality factor.

Indeed, calculating the averages we arrive at
\begin{eqnarray}
& \displaystyle \langle x^2\rangle = \sigma_x^2(t)\, (n+1/2) + \langle u\rangle^2 t^2,\ \langle p^2\rangle = \langle p\rangle^2 + \sigma_p^2\, (n+1/2),
\label{HGavSt}
\end{eqnarray}
and so
\begin{eqnarray}
& \displaystyle \epsilon_x = \lambda_c\,\left(n + \frac{1}{2}\right),\ M = \frac{2\epsilon_x}{\lambda_c} = 2n + 1.
\label{emxSt}
\end{eqnarray}
Thus the quality factor of the orthogonal (standard) HG packet grows linearly with $n$, while for the non-orthogonal (elegant) packet it is only $\sqrt{n}$.

For a packet with the focus at $t=0$ -- that is, $\langle xu\rangle(0) = 0$ -- we have
\begin{eqnarray}
& \displaystyle \sqrt{\langle x^2\rangle - \langle x\rangle^2}|_{t=0}\, \sqrt{\langle p^2\rangle - \langle p\rangle^2} = \frac{\epsilon_x}{\lambda_c} = \frac{M}{2} \approx \left\{
    \begin{array}{ll}
        n,\ \text{standard HG}, \\
        \sqrt{n},\ \text{elegant HG},
    \end{array}
		\right.
\label{emxSt}
\end{eqnarray}
so the quality factor $M$ is just the total number of the mutually orthogonal quantum states in the particle's phase space. Analogously to photons, when the beam consists of many identical wave packets, 
the number $M/2$ defines \textit{information capacity} of the beam. As the quality factor of the non-orthogonal HG packet scales as $\sqrt{n}$, such a beam clearly has a lower information capacity
than the standard one.

Note that the number of states grows with time together with the corresponding entropy $S$ as the packet spreads,
\begin{eqnarray}
& \displaystyle S = \ln\left (\sqrt{\langle x^2\rangle - \langle x\rangle^2}\, \sqrt{\langle p^2\rangle - \langle p\rangle^2}\right ) \propto \ln t.
\label{emxSt}
\end{eqnarray}
This entropy actually is an attribute of a closed system, which includes the packet and an observer, the latter being a macroscopic classical system. 
The observer performed two measurements of $\langle x^2\rangle$ at $t=0$ and then after the time $t$. 
As there were no measurements in between, the growth of the entropy is due to the lost of information by the observer.
However, ``the entropy'' of the packet alone does not grow, which is connected to conservation of the emittance.

\subsubsection{Courant-Snyder formalism}\label{SecCS}

\begin{figure}
\centering
    \includegraphics[width=9cm]{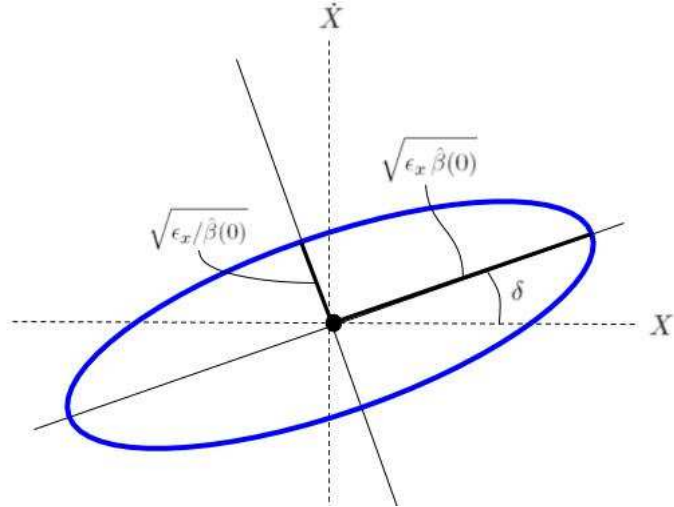}
    \center
		\caption{The phase space of an arbitrary wave packet is described by an ellipse with a minimal area defined by the Schr\"odinger uncertainty relation, $\pi\epsilon_x \geq \pi\lambda_c/2$,
		and by the Twiss or Courant-Snyder parameters (\ref{CS1D2}). When $\delta = 0$, the correlation term $\langle xu\rangle - \langle x\rangle \langle u\rangle$ vanishes and the Schr\"odinger relation reduces to that of Heisenberg. The parameter $\hat{\beta}$ defines evolution of the packet's rms size and the Gouy phase (\ref{phiGSHG}).}
\label{Ellipse}
\end{figure}

An arbitrary wave packet can also be described by \textit{the Courant-Snyder or Twiss} parameters $\hat{\alpha}, \hat{\beta}, \hat{\gamma}$ used in accelerator physics\footnote{Do not confuse $\hat{\beta}$ and $\hat{\gamma}$ with the velocity $\beta$ and the Lorentz factor $\gamma=1/\sqrt{1-\beta^2}$, respectively.} \cite{Reiser}, while the emittance defines a conserved area $\pi\epsilon_x$ of \textit{an ellipse} in the particle's phase space (see Fig.\ref{Ellipse}),
\begin{eqnarray}
& \displaystyle \hat{\beta}\,(\dot{X})^2 + 2\hat{\alpha}\, X\, \dot{X} + \hat{\gamma}\,X^2 = \epsilon_x \geq \frac{\lambda_c}{2},\cr 
& \displaystyle X : = \sqrt{\langle x^2\rangle - \langle x\rangle^2}, \cr
& \displaystyle \text{or}\quad \frac{\hat{\beta}}{2} \left(\langle u^2\rangle - \langle u\rangle^2\right) + \hat{\alpha} \left (\langle xu\rangle - \langle x\rangle \langle u\rangle\right) + \frac{\hat{\gamma}}{2} \left (\langle x^2\rangle - \langle x\rangle^2\right) = \epsilon_x,
\label{CS1D}
\end{eqnarray}
where
\begin{eqnarray}
& \displaystyle\hat{\beta} = \frac{\langle x^2\rangle - \langle x\rangle^2}{\epsilon_x} = \frac{X^2}{\epsilon_x},\quad \hat{\alpha} = -\frac{1}{2}\frac{\partial \hat{\beta}}{\partial t} = -\frac{\langle xu\rangle -\langle x\rangle \langle u\rangle}{\epsilon_x} = -\frac{X\dot{X}}{\epsilon_x},\cr
& \displaystyle \hat{\gamma} = \frac{\langle u^2\rangle - \langle u\rangle^2}{\epsilon_x} = \frac{X\ddot{X} + (\dot{X})^2}{\epsilon_x},\quad \hat{\beta}\hat{\gamma} - \hat{\alpha}^2 = 1.
\label{CS1D2}
\end{eqnarray}
The area of the ellipse is limited from below by the Schr\"odinger uncertainty relation, $\pi\epsilon_x \geq \pi \lambda_c/2$, its slope with respect to a coordinate system is defined 
by the correlation parameter $\hat{\alpha}(\delta)$, and the Twiss parameters at the angle $\delta$ are defined only by $\hat{\beta}(0)$ as follows:
\begin{eqnarray}
& \displaystyle
\begin{pmatrix}
\hat{\alpha}(\delta) \\
\hat{\beta}(\delta) \\ 
\hat{\gamma}(\delta)
\end{pmatrix}
=
\begin{pmatrix}
0 & -\sin\delta\cos\delta & \sin\delta\cos\delta\\
0 & \cos^2\delta & \sin^2\delta\\
0 & \sin^2\delta & \cos^2\delta\\
\end{pmatrix}
\begin{pmatrix}
0 \\
\hat{\beta}(0) \\ 
\hat{\beta}^{-1}(0)
\end{pmatrix},
\label{Ell}
\end{eqnarray}
where we have taken into account that $\hat{\alpha}(0) = 0, \hat{\gamma}(0) = \hat{\beta}^{-1}(0)$.

One can describe the packet's rms size in terms of $\hat{\beta}$-function. For orthogonal packets, the $\hat{\beta}$-function does not depend on the quality factor. For instance, for the standard HG packets
\begin{eqnarray}
& \displaystyle \hat{\beta}^{\text{HG}}(t) = \frac{\sigma^2_x(t)}{\lambda_c} \gg \sigma_x(t),
\label{bHG}
\end{eqnarray}
while for the non-orthogonal packets it can still depend on this factor (say, for the elegant HG packets $\hat{\beta} \propto \sqrt{n},\, n\gg 1$). That is why the use of the Courant-Snyder parameters is more convenient for orthogonal packets. Moreover, the Gouy phase in Eq.(\ref{HGxnSt}) can also be written as follows:
\begin{eqnarray}
& \displaystyle \varphi^{G} = \int \frac{dt}{\hat{\beta}(t)},
\label{phiGSHG}
\end{eqnarray}
while this is \textit{not} the case for the elegant beam (\ref{HGxn}). Thus the Gouy phase of an arbitrary orthogonal packet is analogous to \textit{the Courant-Snyder phase advance} 
employed in classical beam physics \cite{Floettmann2020, Reiser}. Its overal factor $M$ measures the packet's ``quantumness'' according to the Schr\"odinger uncertainty relation. 

\subsection{More than $1$ dimension}
\subsubsection{Hermite-Gaussian packets}

In a $3+1$ dimensional space-time, a wave function of the HG packet simply factorizes,
\begin{eqnarray}
& \displaystyle \psi_{n,j,k}({\bm r}, t) = \psi_n(x,t) \psi_j(y,t) \psi_k(z,t),
\label{HG3D}
\end{eqnarray}
where each of the factors is either the elegant or the standard HG packet. So, the dynamics along each of the coordinates is independent.
Let the packet propagate along the $z$ axis on average, 
$$
\langle{\bm r}\rangle = \langle{\bm u}\rangle t = \{0, 0, \langle u\rangle t\},\ \langle x\rangle = \langle y\rangle = 0.
$$
Then a transverse wave function of the standard HG packet is
\begin{eqnarray}
& \displaystyle \psi_{n,j}(x, y, t) = \psi_n(x,t) \psi_j(y,t) = \frac{N}{(1 + t^2/t_{d,x}^2)^{1/4}(1 + t^2/t_{d,y}^2)^{1/4}}\, H_n\left(\frac{x}{\sigma_x(t)}\right)\, H_j\left(\frac{y}{\sigma_y(t)}\right)\cr 
& \displaystyle \times \exp \Bigg\{-\frac{i}{2}\, (2 n + 1) \arctan\frac{t}{t_{d,x}} - \frac{i}{2}\, (2 j + 1) \arctan\frac{t}{t_{d,y}} - \cr
& \displaystyle \qquad \qquad \qquad - \frac{x^2}{2\sigma_x^2(t)} \left(1 - i\frac{t}{t_{d,x}}\right)
- \frac{y^2}{2\sigma_y^2(t)} \left(1 - i\frac{t}{t_{d,y}}\right)\Bigg\}
\label{HG2D}
\end{eqnarray}
The coefficients at the Gouy phases are simply the packet's quality factors,
\begin{eqnarray}
& \displaystyle M_x = 2n + 1,\ M_y = 2j + 1,
\label{M2D}
\end{eqnarray}
whereas both the Gouy phases themselves can be written as follows:
\begin{eqnarray}
& \displaystyle \arctan \frac{t}{t_{d,x}} = \int \frac{d(t/t_{d,x})}{1 + t^2/t_{d,x}^2} = \int \frac{dt}{\hat{\beta}_{x}(t)},
\label{G2D}
\end{eqnarray}
where we have used the fact that
\begin{eqnarray}
& \displaystyle \hat{\beta}_{x}(t) = \frac{\langle x^2\rangle}{\epsilon_x} = \hat{\beta}_{x}(0)\left(1 + t^2/t_{d,x}^2\right).
\label{beta2D}
\end{eqnarray}

\subsubsection{Laguerre-Gaussian packets}

As known from the works by Allen et al.\cite{Allen}, one can make a packet with a coupling in the transverse plane, 
which is a hallmark of the orbital angular momentum with respect to the propagation axis,
\begin{eqnarray}
& \displaystyle \langle \hat{L}_z\rangle = \int d^3x\, \psi^*({\bm r}, t)\, \hat{L}_z\, \psi({\bm r}, t) = \ell,\ \ell =0, \pm 1,\pm 2,..., \cr
& \displaystyle \int d^3x |\psi({\bm r}, t)|^2 = 1,\ \hat{L}_z = [{\bm r} \times \hat{\bm p}]_z = -i\frac{\partial}{\partial \phi}.
\label{Lzfr}
\end{eqnarray}
For this, one needs to change in the momentum space
$$
\left(p_x - i\, \text{const}\,\frac{\partial}{\partial p_x}\right)^n \left(p_y - i\, \text{const}\,\frac{\partial}{\partial p_y}\right)^j \rightarrow \left(p_x^2 + p_y^2\right)^{|\ell|/2}\, L_{n}^{|\ell|} \left(\frac{p_x^2 + p_y^2}{\sigma_p^2}\right)
$$ 
where $L_{n}^{|\ell|}$ are associated Laguerre polynomials. The corresponding transverse function 
\begin{eqnarray}
& \displaystyle \psi_{\ell, n}({\bm \rho},t) = N \frac{\rho^{|\ell|}}{(\sigma_{\perp}(t))^{|\ell| + 1}}\ L_{n}^{|\ell|}\left(\frac{\rho^2}{(\sigma_{\perp}(t))^2}\right) \exp\Big\{i\ell\phi_r 
- i (2n + |\ell| + 1)\arctan (t/t_d) - \cr
& \displaystyle - \frac{\rho^2}{2(\sigma_{\perp}(t))^2}\, (1-it/t_d)\Big\},\ \int d^2 \rho\, |\psi_{\ell, n}({\bm \rho},t)|^2 = 1,\cr
& \displaystyle \sigma_{\perp}(t) = \sigma_{\perp}(0) \sqrt{1 + \frac{t^2}{t_d^2}},\ \sigma_{\perp}(0) = 1/\sigma_p,
\label{LGSt}
\end{eqnarray}
is an exact non-stationary solution to the 2-dimensional Schr\"odinger equation, called a \textit{standard} Laguerre-Gaussian (LG) packet\footnote{The differences of the LG packet (\ref{LGSt}) 
from those of Refs.\cite{Bliokh, Lloyd} are important for non-relativistic particles \cite{PRA18, PRA19}.}. The set of these functions is complete and orthogonal.

Analogously to the HG packets, there is also an \textit{elegant} Laguerre-Gaussian packet, 
\begin{eqnarray}
& \displaystyle \psi_{\ell, n}({\bm \rho},t) = N \frac{\rho^{|\ell|}}{(\sigma_{\perp}(t))^{n + |\ell| + 1}}\ L_{n}^{|\ell|}\left(\frac{\rho^2}{2 (\sigma_{\perp}(t))^2}(1-it/t_d)\right)\cr
& \displaystyle \times \exp\Big\{i\ell\phi_r - i (n + |\ell| + 1)\arctan (t/t_d) - \frac{\rho^2}{2(\sigma_{\perp}(t))^2}\, (1-it/t_d)\Big\},
\label{LGEl}
\end{eqnarray}
which also satisfies two-dimensional Schr\"odinger equation, has an OAM $\langle \hat{L}_z\rangle = \ell$ with its vanishing uncertainty, but is not orthogonal.
One can also combine LG and HG packets for instance as follows:
\begin{eqnarray}
& \displaystyle \psi_{n,\ell,k}({\bm r}, t) = \psi_{\ell, n}({\bm\rho},t) \psi_k(z,t),
\label{LGHG3D}
\end{eqnarray}
where $\psi_{\ell, n}({\bm\rho},t)$ is either a standard or an elegant LG packet, while the longitudinal function $\psi_k(z,t)$ represents one of the two HG packets.

\subsubsection{Transverse phase space and the Courant-Snyder formalism}

Regardless of the specific model, quantum dynamics of a wave packet in a 3-dimensional space-time is governed by the same Hamiltonian $\hat{H}_{\text{fr}} = \hat{{\bm p}}^2/2m$
and the mean-square radius of the wave packet $\langle\rho^2\rangle$ is not conserved as
\begin{eqnarray}
& \displaystyle \frac{\partial^2\langle\rho^2\rangle(t)}{\partial t^2} = - \left\langle\left[\left[\rho^2,\hat{H}_{\text{fr}}\right],\hat{H}_{\text{fr}}\right]\right\rangle = 2 \langle{\bm u}_{\perp}^2\rangle.
\label{Heiseqfree}
\end{eqnarray}
The general solution to this Heisenberg equation,
\begin{eqnarray}
& \displaystyle \langle\rho^2\rangle(t) = \langle\rho^2\rangle(0) + \frac{\partial\langle\rho^2\rangle(0)}{\partial t}\, t + \langle{\bm u}_{\perp}^2\rangle\, t^2,
\label{rhofree}
\end{eqnarray}
is analogous to Eq.(\ref{xfree}). Note that $\langle\rho^2\rangle(t)$ is connected to an intrinsic \textit{electric quadrupole moment}, which is generally not vanishing for a non-Gaussian packet \cite{Moments, SilenkoQ}. As its third derivative vanishes, the quadrupole moment does not emit electromagnetic waves \cite{LL2}. If the focus is at $t = 0$, the correlation term in (\ref{rhofree}) vanishes and we recover the familiar result,
\begin{eqnarray}
& \displaystyle \langle\rho^2\rangle(t) = \langle\rho^2\rangle(0) + \langle{\bm u}_{\perp}^2\rangle\, t^2 = \langle\rho^2\rangle(0) \left(1 + \frac{t^2}{t_d^2}\right) = \langle\rho^2\rangle(0) \left(1 + \frac{\langle z\rangle^2}{z_R^2}\right),
\label{rhofree2}
\end{eqnarray}
where
\begin{eqnarray}
& \displaystyle t_d = \sqrt{\frac{\langle\rho^2\rangle(0)}{\langle{\bm u}_{\perp}^2\rangle}},\ z_R = \langle u\rangle t_d
\label{td}
\end{eqnarray}
are a diffraction time and a Rayleigh length, respectively. 

Similar to before, one can define a matrix of central moments,
\begin{eqnarray}
& \displaystyle Q_{ij} = \langle {\bm q}_i\cdot {\bm q}_j\rangle - \langle {\bm q}_i\rangle\cdot\langle {\bm q}_j\rangle,\ {\bm q}_i = \{{\bm \rho}, {\bm u}_{\perp}\},\, i=1,2.
\label{Q2D}
\end{eqnarray}
The square root of its determinant, 
\begin{eqnarray}
& \displaystyle \epsilon_{\rho} = \sqrt{\det Q_{ij}} = \sqrt{\langle\rho^2\rangle(t) \langle{\bm u}_{\perp}^2\rangle - \left(\langle{\bm \rho}\cdot{\bm u}_{\perp} \rangle(t)\right)^2} = \sqrt{\langle\rho^2\rangle(0) \langle{\bm u}_{\perp}^2\rangle - \left(\langle{\bm \rho}\cdot{\bm u}_{\perp} \rangle(0)\right)^2},
\label{em}
\end{eqnarray}
is called \textit{a normalized radial emittance} of a packet propagating along the $z$ axis. If the focus is at $t=0$, the result
\begin{eqnarray}
& \displaystyle \epsilon_{\rho} = \sqrt{\langle\rho^2\rangle(0) \,\langle{\bm u}_{\perp}^2\rangle} \geq \lambda_c
\label{emfr3}
\end{eqnarray}
obeys the Heisenberg uncertainty relation. One can define \textit{a transverse quality factor} as follows: 
\begin{eqnarray}
& \displaystyle M = \frac{\epsilon_{\rho}}{\lambda_c} \geq 1,
\label{Mtrarb}
\end{eqnarray}
A packet with $M=1$ minimizes the Schr\"odinger uncertainty relation and it can be called \textit{a fundamental (or Gaussian) mode}. 

So the diffraction time and the Rayleigh length become
\begin{eqnarray}
& \displaystyle t_d = \frac{\langle\rho^2\rangle(0)}{\epsilon_{\rho}} = \frac{\langle\rho^2\rangle(0)}{M \lambda_c},\ z_R = \langle u\rangle t_d.
\label{tdem}
\end{eqnarray}
It might seem that these quantities decrease with the growth of the quality factor $M$. 
If this were the case, the diffraction time could drop below the Compton time $t_c$ (\ref{tc}) for $M \gg 1$. However, for a single-particle state with a stable QED vacuum 
the coordinate uncertainty $\sqrt{\langle\rho^2\rangle(0)}$ must be larger than $\lambda_c$, the momentum uncertainty $\sqrt{\langle{\bm p}_{\perp}^2\rangle(0)}$ 
must be smaller than $m$ for the fundamental mode, and all the time intervals must be larger than the lifetime of a virtual electron-positron pair $t_c$. This implies that 
\begin{eqnarray}
& \displaystyle t_d \geq t_c,\quad \sqrt{\langle\rho^2\rangle(0)} \geq \lambda_c\, M^{(1+\zeta)/2},\ \sqrt{\langle{\bm p}_{\perp}^2\rangle(0)} \leq \lambda_c^{-1}\, M^{(1-\zeta)/2},
\label{rhosc}
\end{eqnarray}
where $\zeta = 0$ corresponds to the generalized coherent states (say, the LG packets), while $\zeta > 0$ corresponds to \textit{a squeezed state} of a particle.
In the former case, the diffraction time does not depend on the packet's quality at all, $t_d = \mathcal O(M^0)$,
while for the squeezed state such a dependence generally exists, $t_d \propto M^{\zeta}$.

It is natural to introduce the following Courant-Snyder variables:
\begin{eqnarray}
& \displaystyle 
\hat{\beta}(t) = \frac{\langle\rho^2\rangle(t)}{\epsilon_{\rho}},\ \hat{\alpha}(t) = -\frac{1}{2}\frac{\partial \beta(t)}{\partial t} = -\frac{\langle{\bm \rho}{\bm u}_{\perp}\rangle(t)}{\epsilon_{\rho}},\ \hat{\gamma} = \frac{\langle{\bm u}_{\perp}^2\rangle}{\epsilon_{\rho}},\cr
& \displaystyle \hat{\beta}\hat{\gamma} - \hat{\alpha}^2 = 1,\ \hat{\beta}(0) = t_d = z_R/\langle u\rangle,
\label{CSel}
\end{eqnarray}
so the conservation of the packet's emittance can be written as an ellipse equation
\begin{eqnarray}
& \displaystyle 
\hat{\gamma}\, P^2 + 2\hat{\alpha}\, P\dot{P} + \hat{\beta}\, (\dot{P})^2 = \epsilon_{\rho} \geq \lambda_c,\cr
& \displaystyle P : = \sqrt{\langle\rho^2\rangle},\cr
& \displaystyle 
\text{or}\quad \frac{\hat{\gamma}}{2}\, \langle\rho^2\rangle + \hat{\alpha}\, \langle{\bm \rho}{\bm u}_{\perp}\rangle + \frac{\hat{\beta}}{2}\, \langle{\bm u}_{\perp}^2\rangle = \epsilon_{\rho} \geq \lambda_c.
\label{CSel2}
\end{eqnarray}
Clearly, for the generalized coherent states with $\zeta = 0$ these functions do not depend on $M$. They are most convenient for an orthogonal set, 
which we imply here, while for a non-orthogonal set (say, the elegant LG packets) the Courant-Snyder parameters can depend on the beam quality factor, which is \textit{not} related to squeezing.
Analogously to Eq.(\ref{G2D}), one can also define a Gouy phase of an arbitrary packet as follows
\begin{eqnarray}
& \displaystyle \varphi^G = \arctan \frac{t}{t_d} = \int \frac{dt}{\hat{\beta}(t)}.
\label{G2DLG}
\end{eqnarray}

We exemplify these definitions with the standard LG packet (\ref{LGSt}) with $\zeta = 0$ for which we find the following averages:
\begin{eqnarray}
& \displaystyle \langle\rho^2\rangle = \sigma_{\perp}^2(t)\, M,\quad \langle{\bm u}_{\perp}^2\rangle = \frac{\sigma_p^2}{m^2}\, M = \frac{\lambda_c^2}{\sigma_{\perp}^2(0)}\, M,\cr
& \displaystyle \epsilon_{\rho} = \lambda_c\,(n + |\ell| + 1) \geq \lambda_c,\quad M = n + |\ell| + 1,\cr
& \displaystyle \hat{\beta} = \frac{\sigma_{\perp}^2(t)}{\lambda_c} =  \frac{\langle\rho^2\rangle(t)_{M=1}}{\lambda_c}.
\label{uperpmean}
\end{eqnarray}

\subsection{Generalized van Cittert–Zernike theorem}\label{SecCZ}

Now we return to the spreading law of an arbitrary quantum packet, Eq.(\ref{rhofree2}), and note that when the spreading is essential -- i.e., the detector is in the \textit{far field} -- we have
\begin{eqnarray}
& \displaystyle \sqrt{\langle\rho^2\rangle(t)} = \sqrt{\langle{\bm u}_{\perp}^2\rangle}\, t = \frac{\langle z\rangle}{\sqrt{\langle\rho^2\rangle(0)}}\, M\,\frac{\lambda_{\text{dB}}}{2\pi},\ \lambda_{\text{dB}} = \frac{2\pi}{\langle p\rangle}.
\label{farf}
\end{eqnarray}
As a result, there is a simple relation between the registered width of the packet at the detector $\sqrt{\langle\rho^2\rangle(\langle z\rangle)}$ and the initial (emitted) width $\sqrt{\langle\rho^2\rangle(0)}$ at the source:
\begin{eqnarray}
& \displaystyle \sqrt{\langle\rho^2\rangle(0)} = \frac{\langle z\rangle \lambda_{\text{dB}}}{2\pi \sqrt{\langle\rho^2\rangle(\langle z\rangle)}}\, M.
\label{theor}
\end{eqnarray}
For the fundamental mode with $M=1$, this is the well-known van Cittert–Zernike theorem for a uniform source \cite{Mandel, Cho, Cho2013, Lat, Ehberger}
where the transverse coherence length $\xi_{\perp} = \sqrt{2}\, \sqrt{\langle\rho^2\rangle(\langle z\rangle)}$ and the effective source radius $r_{\text{eff}} = \sqrt{2}\,\sqrt{\langle\rho^2\rangle(0)}$.
Eq.(\ref{theor}) is also applicable for a ``more quantum'' packet with $M > 1$, that is, not Gaussian one.

Moreover, the general spreading law (\ref{rhofree2}) is applicable not only in the far field, but also nearby the source. 
In particular, one can keep the first correction to the van Cittert–Zernike theorem when the detector is in \textit{a Fresnel zone},
\begin{eqnarray}
& \displaystyle \sqrt{\langle\rho^2\rangle(\langle z\rangle)} = \frac{\langle z\rangle \lambda_{\text{dB}}}{2\pi \sqrt{\langle\rho^2\rangle(0)}}\, M\, \left (1 + \frac{1}{2}\left(\frac{2\pi \langle\rho^2\rangle(0)}{\langle z\rangle \lambda_{\text{dB}}}\frac{1}{M}\right)^2\right),
\label{CZth}
\end{eqnarray}
where the second term in the r.h.s. vanishes in the far field when (see Eq.(\ref{tdem}))
\begin{eqnarray}
& \displaystyle \langle z\rangle \gg z_R = \frac{2\pi}{M}\frac{\langle\rho^2\rangle(0)}{\lambda_{\text{dB}}}.
\label{CZth2}
\end{eqnarray}
For electrons with $\sqrt{\langle\rho^2\rangle} \sim 1\, \text{nm},\ \beta \sim 10^{-3}-10^{-1}$, the Rayleigh length $z_R$ lies between $1$ nm and $100$ nm, see Eq.(\ref{Raylel}) below.

Here a note on coherence is in order. Although Eq.(\ref{rhofree2}) was derived quantum-mechanically for a single-particle packet, the very same spreading law takes place for a classical beam of many point particles \cite{Reiser, Floettmann2020} where $\langle ...\rangle$ is an average over a particle distribution in the beam\footnote{As a result, the quality factor of the classical beam is conserved in linear fields.}. In an accelerator, such a beam usually represents an \textit{incoherent} mixture of particles and the wave packets do not interfere, in accord with the fact that the van Cittert–Zernike theorem implies an incoherent source. Remarkably, the quantum derivation of Eq.(\ref{rhofree2}) explicitly shows that the packets inside the beam or the plane-wave components of a single packet \textit{can} interfere coherently. Moreover, the packet of a massive particle \textit{is fully coherent} with itself when $\sqrt{\langle\rho^2\rangle}\gg \lambda_c$. This coherence can be violated for such tightly focused quantum superpositions with $\sqrt{\langle\rho^2\rangle}\gtrsim \lambda_c$ as, for instance, the Schr\"odinger cat states with a not-everywhere positive Wigner function \cite{Case}. Thus the generalized van Cittert–Zernike theorem (\ref{theor}) or (\ref{CZth}) holds valid both for coherent superpositions of quantum states with a possible fermionic repulsion (e.g., for electrons emitted from cold cathodes and for beams of electron microscopes) and for incoherent mixtures such as accelerator beams. Finally, note that semi-classically one can derive the far-field van Cittert–Zernike theorem (\ref{theor}) and its $M$ scaling from a Bragg condition of constructive interference of two waves in a beam,
that is, implying full coherence.

\section{Quantum dynamics in electric and magnetic lenses} 

\subsection{In a solenoid}\label{SecSol}

We start with the constant and homogeneous magnetic field (\ref{Adef}) and define an operator of the canonical momentum 
\begin{eqnarray}
& \displaystyle
\hat{\bm p}^{\text{can}} = \hat {\bm p}^{\text{kin}} + e {\bm A} = -i{\bm\nabla}.
\label{oop}
\end{eqnarray}
There exist two angular momenta,
\begin{eqnarray}
& \displaystyle \hat{{\bm L}}^{\text{can}} ={\bm r} \times \hat{{\bm p}}^{\text{can}}\quad \text{and}\quad \hat{{\bm L}}^{\text{kin}} ={\bm r} \times \hat{{\bm p}}^{\text{kin}}.
\label{Lkin}
\end{eqnarray}
The $z$-component of the former, $\hat{L}_z^{\text{can}} = -i\partial/\partial \phi$, commutes with the Hamiltonian
\begin{eqnarray}
& \displaystyle \hat{H} = \frac{(\hat{\bm p}^{\text{kin}})^2}{2m} = \frac{(\hat{\bm p}^{\text{can}})^2}{2m} -\omega_L \hat{L}_z^{\text{can}} + \frac{m}{2}\,\omega_L^2\, \rho^2
\label{H}
\end{eqnarray}
and therefore $\langle \hat{L}_z^{\text{can}}\rangle$ is conserved in the field, 
exactly like for the free twisted particle. Let the packet propagate along the z-axis on average, 
\begin{eqnarray}
& \displaystyle \langle {\bm r}\rangle(t) =\{0, 0, \langle u\rangle t\},\ \langle{\bm u}\rangle = \frac{\langle\hat{\bm p}^{\text{kin}}\rangle}{m},\cr
& \displaystyle \langle\hat{\bm p}^{\text{kin}}_{\perp}\rangle = \langle\hat{{\bm p}}_{\perp}^{\text{can}}\rangle = 0,\ 
\langle\hat p^{\text{kin}}_z\rangle = \langle\hat p^{\text{can}}_z\rangle =  \langle u\rangle\, m,
\label{rmean}
\end{eqnarray}
and be twisted,
\begin{eqnarray}
& \displaystyle \langle \hat{L}_z^{\text{can}}\rangle = \ell,\ \ell =0, \pm 1,\pm 2,...
\label{Lz}
\end{eqnarray}

While we start with the field which is \textit{homogeneous in all space}, we will later consider the particle entering a solenoid of a finite length at a certain moment of time. 
As the initial transverse momentum is vanishing, the packet's centroid does not rotate and the mean path stays rectilinear.
The canonical angular momentum of the incoming free packet is not vanishing and its conservation in such a field means that if a twisted particle enters a solenoid, 
moves along the field lines, and then leaves it, its final OAM will stay the same. In other words, vortex particles \textit{can be trapped and focused} 
by the solenoids and magnetic lenses without deteriorating their quantum states. However, any particle that is off the solenoid axis will get a transverse kick when entering a magnetic lens, so the mean kinetic momentum will no longer be parallel to the quantization axis of the canonical AM (z). That is why for a vortex beam of many particles a thick lens will inevitably lead to broadening of an OAM spectrum (an OAM bandwidth), eventhough the mean OAM is conserved.

Unlike $\langle \hat{L}_z^{\text{can}}\rangle$, its kinetic counterpart,
\begin{eqnarray}
& \displaystyle \langle \hat{L}_z^{\text{kin}}\rangle = \ell -m\omega_L \langle\rho^2\rangle = \ell - 2\,\, \sgn(e) \frac{\langle\rho^2\rangle}{\rho_H^2},
\label{Lzkin}
\end{eqnarray}
is \textit{not} generally conserved in time, as $\rho^2$ does not commute with the Hamiltonian. 
Here
\begin{eqnarray}
& \displaystyle \rho_H = \sqrt{\frac{4}{|e| H}} = 2\lambda_c \sqrt{\frac{H_c}{H}}
\label{rH}
\end{eqnarray}
is denoted (recall that $H > 0$ and $H_c$ is from Eq.(\ref{Hc})). For typical field strengths of $H \sim 0.1$\,T -- $10$\,T, we have
\begin{eqnarray}
& \displaystyle \rho_H \sim 10\, \text{nm} - 100\, \text{nm}.
\label{rhoH2}
\end{eqnarray} 

There exists an exact stationary solution to the Schr\"odinger equation in the given field, which satisfies the above requirements -- the so-called Landau states \cite{LL3, ST, Bliokh12},
\begin{eqnarray}
& \displaystyle \Psi ({\bm r}, t) = \text{const}\, \left(\frac{\rho}{\rho_H}\right)^{|\ell|} L_{n_H}^{|\ell|}\left(\frac{2\rho^2}{\rho_H^2}\right) 
\exp\left\{-\frac{\rho^2}{\rho_H^2} + i\ell\phi + i\, p_z z - i\, \varepsilon t\right\},
\label{Dau}
\end{eqnarray}
where $L_{n_H}^{|\ell|}$ are the generalized Laguerre polynomials with $n_H = 0, 1, 2, ...$, and
\begin{eqnarray}
& \displaystyle
\ \varepsilon = \varepsilon_{\perp} + \frac{p_z^2}{2m},\cr
& \displaystyle \varepsilon_{\perp} = \frac{\left\langle(\hat{\bm p}_{\perp}^{\text{kin}})^2\right\rangle}{2m} = |\omega_L| 
\left(2n_H + |\ell| - \sgn(e)\,\ell + 1\right) \geq |\omega_L|.
\label{epsilon}
\end{eqnarray}
The mean-square radius $\langle\rho^2\rangle_{\text{LG}}$ calculated with these states is
\begin{eqnarray}
& \displaystyle \langle\rho^2\rangle_{\text{LG}} = \frac{\rho_H^2}{2}\, (2n_H + |\ell| + 1) \geq \frac{\rho_H^2}{2}.
\label{rho2LG}
\end{eqnarray}
As it is independent of time, the kinetic AM 
\begin{eqnarray}
& \displaystyle \langle \hat{L}_z^{\text{kin}}\rangle = \ell - \sgn(e)\, (2n_H + |\ell| + 1)
\label{LzkinLG}
\end{eqnarray}
\textit{is also conserved} and its minimum value is $\langle \hat{L}_z^{\text{kin}}\rangle_{\text{min}} = - \sgn(e)$.
However, this conclusion only holds in the stationary (thick-lens) regime -- that is, if the time $\Delta t$ during which the particle stays in the solenoid is at least one rotation period,
\begin{eqnarray}
& \displaystyle \Delta t = T_c = \frac{2\pi}{\omega_c}\ \text{or} \ \Delta t \gg T_c.
\label{dtin}
\end{eqnarray}
So the stationary Landau states are not applicable for describing quantum dynamics of the wave packet in \textit{a thin-lens regime} when $\Delta t \ll T_c$.

To describe this dynamics for arbitrary $\Delta t$, we need to solve the Heisenberg equations,
\begin{eqnarray}
& \displaystyle \frac{\partial^2\langle\rho^2\rangle(t)}{\partial t^2} = - \left\langle\left[\left[\rho^2,\hat{H}\right],\hat{H}\right]\right\rangle = \cr
& \displaystyle = -\omega_c^2\, \left(\langle\rho^2\rangle(t) - \langle\rho^2\rangle_{\text{st}}\right),
\label{Heiseq}
\end{eqnarray}
where
\begin{eqnarray}
& \displaystyle \langle\rho^2\rangle_{\text{st}} = \frac{1}{m\omega_L} \left(\frac{\varepsilon_{\perp}}{\omega_L} + \ell \right).
\label{r0}
\end{eqnarray}
The general solution to this equation contains two integration constants $\alpha_1$ and $\alpha_2$,
\begin{eqnarray}
& \displaystyle \langle\rho^2\rangle(t) = \langle\rho^2\rangle_{\text{st}} + \alpha_1 \cos\left(\omega_c t + \alpha_2\right).
\label{r0gen}
\end{eqnarray}
Thus although the packet's centroid moves rectilinearly, the mean-square radius $\langle\rho^2\rangle$ oscillates with the period of $T_c$.
This periodic behaviour is sometimes called \textit{quantum broadening of the classical trajectories} \cite{ST, STJ}.
Note that this rms radius defines an intrinsic electric quadrupole moment of the packet \cite{Moments, SilenkoQ}. 
As its third derivative does not vanish, this packet emits electromagnetic waves \cite{LL2}.
Averaging Eq.(\ref{r0gen}) over at least one period, we return to the stationary (thick lens) regime with 
\begin{eqnarray}
& \displaystyle
\overline{\left\langle\rho^2\right\rangle(t)} = \langle\rho^2\rangle_{\text{st}} = \langle\rho^2\rangle_{\text{LG}},
\label{avrho}
\end{eqnarray}
for which the Landau states are applicable. 

The equations of motion for $\rho^2$ can also be obtained by the Ehrenfest theorem,
\begin{eqnarray}
& \displaystyle \frac{d\langle{\bm \rho}\rangle}{dt} = \langle{\bm u}_{\perp}\rangle,\ \frac{d^2\langle{\bm \rho}\rangle}{dt^2} = \frac{e}{m}\, \left\langle{\bm u} \times {\bm H}\right\rangle,
\label{eqmotr}
\end{eqnarray}
and so
\begin{eqnarray}
& \displaystyle \frac{\partial^2 \langle\rho^2\rangle (t)}{\partial t^2} = 2\langle {\bm u}_{\perp}^2\rangle + 2\frac{e}{m}\,\langle{\bm \rho}\cdot [{\bm u} \times {\bm H}] \rangle
= 2\langle {\bm u}_{\perp}^2\rangle + 2 \frac{\omega_c}{m}\, \langle \hat{L}_z^{\text{kin}}\rangle (t).
\label{eqmotr2}
\end{eqnarray}
One can rewrite Eq.(\ref{Heiseq}) as follows:
\begin{eqnarray}
& \displaystyle \frac{\partial^2 \langle\rho^2\rangle (t)}{\partial t^2} + 2 K(t) \langle \rho^2\rangle(t) = 2\langle {\bm u}_{\perp}^2\rangle,\cr
& \displaystyle K(t) = -\frac{1}{\langle \rho^2\rangle(t)}\, \left\langle{\bm \rho} \frac{d^2{\bm \rho}}{dt^2}\right\rangle = -\frac{\omega_c}{m} \frac{\langle \hat{L}_z^{\text{kin}}\rangle(t)}{\langle \rho^2\rangle(t)}.
\label{K}
\end{eqnarray}
Taking $\langle \hat{L}_z^{\text{kin}}\rangle(t)$ from Eq.(\ref{Lzkin}), we recover the r.h.s of Eq.(\ref{Heiseq}).
When $\langle \hat{L}_z^{\text{kin}}\rangle(t) \rightarrow 0$ this equation coincides with Eq.(\ref{Heiseqfree}) for the free packet.
However, Eq.(\ref{r0gen}) does not continously transform to Eq.(\ref{rhofree}) when $H \rightarrow 0$ \textit{before the boundary conditions are applied}.

Importantly, Eq.(\ref{Heiseq}) and the conservation of $\langle \hat{L}_z^{\text{can}} \rangle$ also hold for a solenoid of a finite length $L$, 
which is described by the potential 
\begin{eqnarray}
& \displaystyle
{\bm A} = \frac{H(z)}{2}\{-y,x,0\},
\label{ls}
\end{eqnarray}
where $H(z) = H$ when $0 \leq z \leq L$ and zero otherwise (a hard-edge approximation). Such a field has a linear radial component 
\begin{eqnarray}
& \displaystyle
H_{\rho}(\rho,z) = -\frac{\rho}{2}\,H^{\prime}(z),\ H^{\prime}(z) \equiv \frac{\partial H(z)}{\partial z},
\label{Hrho}
\end{eqnarray} 
which vanishes at the solenoid axis and leads to \textit{focusing and rotation} of a particle that is off this axis. 

Moreover, although the field of a real solenoid is \textit{not} radially homogeneous, its dependence on $\rho$ can be neglected, $H(\rho,z) \approx H(0,z) \equiv H(z)$, 
when the field does not change over the packet's transverse coherence length, which is 
\begin{eqnarray}
& \displaystyle
\sqrt{\langle\rho^2\rangle_{\text{LG}}} \sim \rho_H \sqrt{2n_H + |\ell| + 1} \sim (10\,\text{nm} - 100\,\text{nm})\,\sqrt{2n_H + |\ell| + 1}
\label{restim}
\end{eqnarray}
Clearly, the radius $\sqrt{\langle\rho^2\rangle_{\text{LG}}}$ is usually much smaller than the typical widths of the multi-particle beams, which are of the order of $10\,\mu\text{m} - 10$\,mm. So while the approximation of a radially homogeneous field may fail for the beam as a whole, it can still hold for each individual packet. Indeed, for the \textit{inhomogeneous} field we have instead of Eq.(\ref{Lzkin}):
\begin{eqnarray}
& \displaystyle \langle \hat{L}_z^{\text{kin}}\rangle = \ell - e \langle\rho A_{\phi}\rangle =  \ell - \frac{e}{2} \langle\rho^2 H(\rho,z)\rangle,\cr
& \displaystyle \langle\rho^2 H(\rho,z)\rangle = \int d^3r\, \Psi^*({\bm r}, t)\,\rho^2 H(\rho,z)\, \Psi({\bm r}, t),
\label{Lzkin2}
\end{eqnarray}
where $\Psi({\bm r}, t) \propto \exp\{-\rho^2/\rho_H^2\}$ are the Landau states (\ref{Dau}). 
Therefore, one can put $H(\rho,z) \approx H(z)$ when it is nearly constant over the distance of $\rho_H$.
The solenoid fields can generally be presented in a form of series \cite{Reiser},
\begin{eqnarray}
& \displaystyle
H_z(\rho,z) = H_z(0,z) - \frac{\rho^2}{4}\,H_z^{\prime\prime}(0,z) + \mathcal O(\rho^4),\cr
& \displaystyle
H_{\rho}(\rho,z) = -\frac{\rho}{2}\,H^{\prime}(z) + \mathcal O(\rho^3),
\label{HSolS}
\end{eqnarray}
and so the non-linear terms are generally \textit{less important} for single quantum wave packets than for wide beams of accelerators.

\subsection{Entering the solenoid}\label{SecEnSol}

Let the standard LG packet (\ref{LGSt}) enter the solenoid at $t=0, \langle z\rangle=0$. The central moments in the transverse plane,
\begin{eqnarray}
& \displaystyle Q_{ij}(t) = \langle {\bm q}_i\cdot{\bm q}_j \rangle - \langle {\bm q}_i\rangle\cdot\langle{\bm q}_j \rangle,\ {\bm q}_i = \{{\bm \rho}, {\bm u}_{\perp}\},\ i=1,2,
\label{cmomr}
\end{eqnarray}
are continuous at the boundary together with the emittance $\epsilon_{\rho} = \sqrt{\det Q_{ij}}$ and with the packet's transverse quality factor $M = \epsilon_{\rho}/\lambda_c$,
\begin{eqnarray}
& \displaystyle Q_{ij}^{\text{free}}(0) = Q_{ij}^{\text{field}}(0),\ \epsilon_{\rho}^{\text{free}} = \epsilon_{\rho}^{\text{field}}(0),\ M^{\text{free}} = M^{\text{field}}(0).
\label{cmomr2}
\end{eqnarray}
This implies the boundary conditions
\begin{eqnarray}
& \displaystyle \langle\rho^2\rangle^{\text{free}}(0) = \langle\rho^2\rangle^{\text{field}}(0),\, \langle{\bm\rho}{\bm u}_{\perp}\rangle^{\text{free}}(0) = \langle{\bm\rho}{\bm u}_{\perp}\rangle^{\text{field}}(0),\, \langle{\bm u}_{\perp}^2\rangle^{\text{free}}(0) = \langle{\bm u}_{\perp}^2\rangle^{\text{field}}(0).
\label{incond}
\end{eqnarray}
Finding $\alpha_1, \alpha_2$ from these conditions, the rms-radius (\ref{r0gen}) takes the following form (cf. \cite{FA1, FA2}):
\begin{eqnarray}
& \displaystyle \langle\rho^2\rangle(t) = \langle\rho^2\rangle_{\text{st}} + \left(\langle\rho^2\rangle^{\text{free}} - \langle\rho^2\rangle_{\text{st}}\right) \cos\left(\omega_c t\right),
\label{r0ent}
\end{eqnarray}
where $\langle\rho^2\rangle^{\text{free}} \equiv \langle\rho^2\rangle^{\text{free}}(0)$. Now when we switch off the field, $H \to 0$, this solution continuously transforms to $\langle\rho^2\rangle^{\text{free}}$. As $\langle\rho^2\rangle(t)$ must be positive, we have the following inequalities:
\begin{eqnarray}
& \displaystyle \langle\rho^2\rangle_{\text{st}} = \langle\rho^2\rangle_{\text{LG}} > \frac{\langle\rho^2\rangle^{\text{free}}}{2},\cr
& \displaystyle
 \epsilon_{\rho}^{\text{LG}} = \sqrt{\det\, \overline {Q_{ij}(t)}} = \sqrt{\langle\rho^2\rangle_{\text{LG}}\,\langle{\bm u}_{\perp}^2\rangle} > \frac{\epsilon_{\rho}^{\text{free}}}{\sqrt{2}} \geq \frac{\lambda_c}{\sqrt{2}}.
\label{Ineqgen}
\end{eqnarray}
The last condition in Eq.(\ref{incond}) connects the principal quantum number $n_H$ in the magnetic field and the radial quantum number $n$ of the free LG packet,
\begin{eqnarray}
& \displaystyle \sqrt{\frac{2n_H + |\ell| - \sgn(e)\,\ell + 1}{n + |\ell| + 1}} = \frac{1}{2}\, \frac{\rho_H}{\sigma_{\perp}(0)},
\label{nH}
\end{eqnarray}
and therefore when both $n$ and $n_H$ are not vanishing
\begin{eqnarray}
& \displaystyle \frac{n_H}{n} \sim \left(\frac{\rho_H}{\sigma_{\perp}(0)}\right)^2.
\label{nH2}
\end{eqnarray}
When the distance from the solenoid to the electron source is shorter than the Rayleigh length (say, for relativistic electrons), 
the width of the fundamental mode $\sigma_{\perp}$ does not exceed a few nanometers \cite{Cho, Ehberger, Lat} and so $\sigma_{\perp} \ll \rho_{H},\, n_H \gg n$.
The opposite regime is realized when the packet is non-relativistic and it can spread so that\footnote{For a beam of an electron microscope, $\sigma_{\perp}$ can exceed 1 mm.} $\sigma_{\perp} \gg \rho_{H},\, n_H \ll n$.

\begin{figure}
\centering
    \includegraphics[width=16cm]{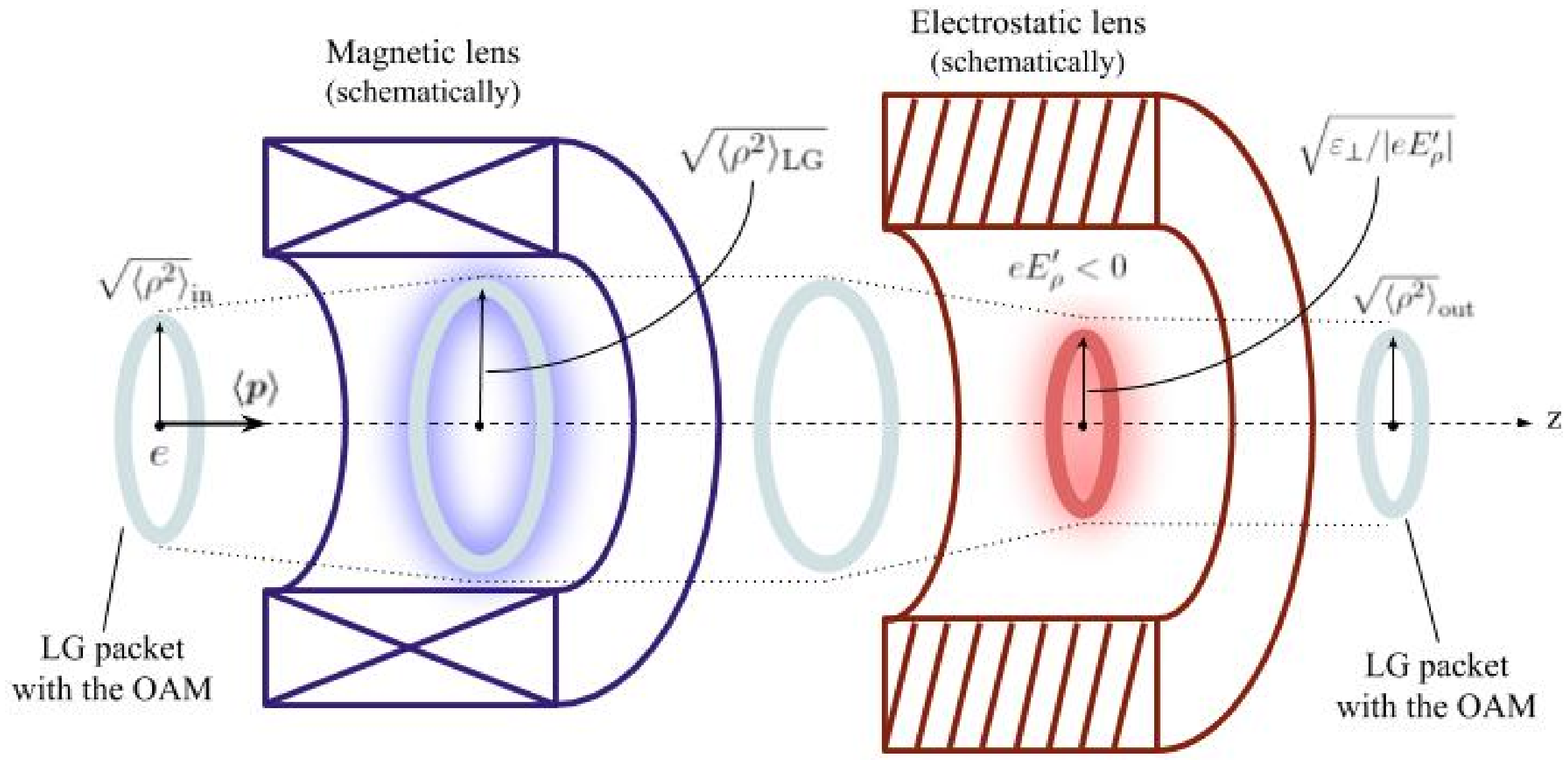}
    \center
		\caption{When transmitted through a combination of linear electric and magnetic lenses, the OAM of a vortex particle stays the same, although its rms radius $\sqrt{\langle\rho^2\rangle}$, the rms velocity $\sqrt{\langle{\bm u}_{\perp}^2\rangle}$, and the emittance may alter. The quantum oscillations (\ref{r0ent}), (\ref{rhoE2st}), and (\ref{rhoHE2st}) blur the rms radius inside the lenses 
		and the final value $\sqrt{\langle\rho^2\rangle}_{\text{out}}$ depends on a phase at which the particle leaves the field.}
\label{Transm}
\end{figure}

The function $\langle\rho^2\rangle(t)$ from Eq.(\ref{r0ent}) has two extrema: at $t = 0$ and at $t = T_c/2 = \pi/\omega_c$. However, which of them is the maximum and which is the minimum depends on whether $\langle\rho^2\rangle_{\text{st}} > \langle\rho^2\rangle^{\text{free}}$ or $\langle\rho^2\rangle_{\text{st}} < \langle\rho^2\rangle^{\text{free}}$. The former case corresponds to $\rho_H > \sigma_{\perp}, n_H > n$ and the lens can be called \textit{defocusing at short times} (shown in Fig.\ref{Transm}), while the latter regime is realized when $\rho_H < \sigma_{\perp}, n_H < n$, that is, the lens is \textit{focusing at short times} and we have 
\begin{eqnarray}
& \displaystyle
\langle\rho^2\rangle^{\text{free}}/2 < \langle\rho^2\rangle_{\text{st}} < \langle\rho^2\rangle^{\text{free}}
\label{InEq}
\end{eqnarray} 
due to (\ref{Ineqgen}). In other words, the rms-radius and the emittance are \textit{nearly constant} for the focusing lens. At large times, $\langle\rho^2\rangle(t)$ oscillates with a period $T_c$ around $\langle\rho^2\rangle_{\text{LG}}$ together with the emittance $\epsilon_{\rho}(t) = \sqrt{\det Q_{ij}(t)}$ with its mean value 
\begin{eqnarray}
& \displaystyle
\epsilon_{\rho}^{\text{LG}} = \sqrt{\det\,\overline {Q_{ij}(t)}} = \sqrt{\langle\rho^2\rangle_{\text{LG}}\,\langle{\bm u}_{\perp}^2\rangle} = \cr
& \displaystyle = \lambda_c\, \sqrt{2}\, \sqrt{(2n_H + |\ell| + 1) (2n_H + |\ell| -\sgn(e) \ell + 1)}.
\label{emmeanH}
\end{eqnarray}
The packet's \textit{effective} quality factor $M = \epsilon_{\rho}^{\text{LG}}/\lambda_c$ can be an irrational number and it can be much larger than the quality factor of the incoming LG packet $M = n + |\ell| + 1$ when $n_H \gg n$, i.e. for the defocusing-at-short-times lens. In this case, the mean emittance in the field $\epsilon_{\rho}^{\text{LG}}$ is larger than $\epsilon_{\rho}^{\text{free}}$, so the packet becomes ``more quantum''\footnote{In contrast to classical beams, larger emittances and larger quality factors of quantum packets are the hallmarks of non-classicality and non-Gaussianity, which is more attractive for applications than the fundamental mode with the minimal emittance.}. In contrast to the free LG packet, the mean emittance and the quality factor depend on the sign of the OAM $\ell$ and of the particle's charge. In particular, when $\sgn(e)\,\ell < 0$ the emittance grows as $\epsilon_{\rho}^{\text{LG}} \propto |\ell|$ for large $\ell$, while it is only  $\epsilon_{\rho}^{\text{LG}} \propto \sqrt{|\ell|}$ for $\sgn(e)\,\ell > 0$. In other words, the electron state with $\ell > 0$ (along the magnetic field) has \textit{a higher quality factor} than the electron state with $\ell < 0$ (opposite to the magnetic field).

In a thin-lens regime with $\Delta t \ll T_c$, we can represent (\ref{r0ent}) in a form, similar to that of a free packet:
\begin{eqnarray}
& \displaystyle \langle\rho^2\rangle(t) \approx \langle\rho^2\rangle^{\text{free}} + \frac{\omega_c^2}{2}\,\left(\langle\rho^2\rangle_{\text{st}} - \langle\rho^2\rangle^{\text{free}}\right) t^2 = \cr
& \displaystyle = \langle\rho^2\rangle^{\text{free}} \left(1 \pm \frac{t^2}{t_d^2}\right) \equiv \langle\rho^2\rangle^{\text{free}} \left(1 \pm \frac{\langle z\rangle^2}{z_R^2}\right),\ z_R = \langle u \rangle\, t_d,
\label{r0entsm}
\end{eqnarray}
where the upper sign corresponds to $\langle\rho^2\rangle_{\text{st}} > \langle\rho^2\rangle^{\text{free}}$ (defocusing) and the lower sign -- to the opposite case (focusing). The diffraction time is
\begin{eqnarray}
& \displaystyle t_d = \frac{T_c}{\pi\sqrt{2}}\,\sqrt{\frac{\langle\rho^2\rangle^{\text{free}}}{|\langle\rho^2\rangle_{\text{st}} - \langle\rho^2\rangle^{\text{free}}|}}
\label{tdH}
\end{eqnarray}
Accordingly, one can define a Lorentz invariant \textit{Gouy phase} and a $\hat{\beta}$-function of the electron packet in a thin magnetic lens:
\begin{eqnarray}
& \displaystyle \varphi^{G} = \pm \arctan \frac{t}{t_d} = \pm \int \frac{dt}{\hat{\beta}(t)} \approx \pm \frac{t}{t_d}.
\label{GH}
\end{eqnarray}
One can also derive this phase from a wave function, which is a solution to the paraxial wave equation in the magnetic field \cite{Silenko2020}.
We see that the condition of paraxiality, 
\begin{eqnarray}
& \displaystyle \sqrt{\langle ({\bm p}^{\text{kin}}_{\perp})^2\rangle} \ll \langle \hat{p}_z^{\text{kin}}\rangle
\label{parcond}
\end{eqnarray}
is intimately connected with the thin-lens condition, $\Delta t \ll T_c$. Both of them imply a non-vanishing longitudinal momentum and the finite time of flight through the field.
These conditions may not be met in a thick lens or in a storage ring. 

Let us estimate how realistic the thin-lens (paraxial) approximation is. The cyclotron period $T_c = 2\pi/\omega_c$ can be presented as follows: 
\begin{eqnarray}
& \displaystyle T_c = 2\pi\, t_c\, \frac{H_c}{H},
\label{TcHc}
\end{eqnarray}
where $t_c$ is the Compton time (\ref{tc}). For the fields strenghts of $H \sim 0.1$\,T -- $10$\,T, we have
\begin{eqnarray}
& \displaystyle T_c \sim 10^{-12}\,\text{sec}. - 10^{-10}\,\text{sec}.
\label{TcHcqt}
\end{eqnarray}
In the laboratory frame of reference this time is $\gamma = 1/\sqrt{1 - \beta^2}$ times larger. The time of flight $\Delta t = L/\langle u\rangle$ through the lens of the length $L \sim 1$\,cm -- $1$\,m is 
\begin{eqnarray}
& \displaystyle \Delta t \sim 10^{-10}\, \text{sec}. - 10^{-8}\, \text{sec}.
\label{dtest}
\end{eqnarray}
We see that the paraxial regime $\Delta t \ll T_c$ can be realized for ultrarelativistic electrons with $\gamma \sim 10^2-10^3$, the thin lens of $L < 10$\, cm, and not very high fields strengths, $H \sim 0.1$ T. A more realistic scenario with $H > 0.1$ T, $\gamma \sim 1 - 10$, $L > 10$ cm is \textit{not} compatible with the paraxial approximation and in this (thick-lens) regime 
the Gouy phase and the $\hat{\beta}$-function (\ref{GH}) do not make sense because the packet \textit{does not spread} on average. 

Now we return to the kinetic AM (\ref{Lzkin}), which is
\begin{eqnarray}
& \displaystyle \langle \hat{L}_z^{\text{kin}}\rangle(t) = \ell - m\omega_L \left(\langle\rho^2\rangle_{\text{st}} + \left(\langle\rho^2\rangle^{\text{free}} - \langle\rho^2\rangle_{\text{st}}\right) \cos\left(\omega_c t\right) \right) = \cr
& \displaystyle = -\frac{\varepsilon_{\perp}}{\omega_L} - m\omega_L \left( \langle\rho^2\rangle^{\text{free}} - \langle\rho^2\rangle_{\text{st}}\right) \cos\left(\omega_c t\right).
\label{Lzkinent}
\end{eqnarray}
In the hard-edge approximation, it experiences a sudden change when the particle enters the field at $t = 0$,
\begin{eqnarray}
& \displaystyle \langle \hat{L}^{\text{kin}}_z\rangle(0) - \langle \hat{L}_z\rangle^{\text{free}} = -m\omega_L \langle\rho^2\rangle^{\text{free}} = -2\,\sgn(e)\, \frac{\langle\rho^2\rangle^{\text{free}}}{\rho_H^2}.
\label{Lzkindelta}
\end{eqnarray}
If the incoming particle is a Gaussian packet with $\langle \hat{L}_z\rangle^{\text{free}} = \ell = 0$, then Eq.(\ref{Lzkindelta}) yields the kinetic AM that the packet acquires when entering the solenoid. In full analogy to a classical beam \cite{Reiser}, all \textit{the Bohmian trajectories} of the packet that are off the axis start to rotate resulting in \textit{a finite azimuthal component} of the current density.

The sudden change of the kinetic AM can also be represented as a quantum average of an Aharonov-Bohm phase,
\begin{eqnarray}
& \displaystyle \left\langle e \oint\frac{d{\bm\rho}}{2\pi}\, {\bm A} \right\rangle = \langle \hat{L}^{\text{kin}}_z\rangle(0) - \langle \hat{L}_z\rangle^{\text{free}}.
\label{deltaLzL}
\end{eqnarray}
Importantly, as the particle's rectilinear path is not changed by the field on the solenoid axis, the classical trajectory is not closed and the flux of the field 
is calculated through \textit{the transverse area of the wave-packet} itself. In other words, the integral is taken not over the cyclotron radius but over all the Bohmian trajectories 
with a subsequent quantum averaging. The corresponding mean flux through the packet's area
\begin{eqnarray}
& \displaystyle \langle\Phi\rangle =  \pi\,H\,\langle\rho^2\rangle^{\text{free}} = - \left\langle \oint d{\bm\rho}\, {\bm A} \right\rangle
\label{Phimean}
\end{eqnarray}
is quantized when the incoming state is \textit{not} the fundamental mode -- for instance, the standard LG packet with $M > 1$,
\begin{eqnarray}
& \displaystyle \langle\Phi\rangle = \pi\,H\,\langle\rho^2\rangle^{\text{free}} = \pi\,H\,\sigma_{\perp}^2\, (n + |\ell| + 1).
\label{Phimean2}
\end{eqnarray}

One can generalize these results when the incoming transverse momentum $\langle\hat{{\bm p}}_{\perp}^{\text{kin}}\rangle$ and the cyclotron radius are not vanishing.
Then the AM (\ref{Lzkin}) contains two contributions: \textit{(i)} $L_z^{\text{cyclo}}$ comes from the cyclotron motion of the packet's centroid 
and \textit{(ii)} $L_z^{\text{w-p}}$ is purely quantum and connected with the current circulating around this centroid:
\begin{eqnarray}
& \displaystyle \langle \hat{L}_z^{\text{kin}}\rangle = L_z^{\text{cyclo}} + L_z^{\text{w-p}},\quad L_z^{\text{cyclo}} = -m\omega_L \langle{\bm \rho}\rangle^2,\cr
& \displaystyle L_z^{\text{w-p}} = \ell - m\omega_L \left(\langle\rho^2\rangle - \langle{\bm \rho}\rangle^2\right) \equiv \langle \hat{L}_z^{\text{can}}\rangle + L_z^{\text{dia}}.
\label{Lzcyclo}
\end{eqnarray}
A part of the latter, $L_z^{\text{dia}}(t) = - m\omega_L \left(\langle\rho^2\rangle(t) - \langle{\bm \rho}\rangle^2\right)$, is sometimes called the diamagnetic AM \cite{FA2}. 
The mean path $\langle {\bm r}\rangle(t)$ coincides with Eq.(\ref{rp}) and so (cf. Eq.(\ref{Lzint})) $\langle{\bm \rho}\rangle^2 = \rho_c^2,\ L_z^{\text{cyclo}} = L_z^{\text{int}}/2$.
In this case $\overline{\langle\rho^2\rangle(t)}$ is to be calculated over the states with $\langle\hat{{\bm p}}_{\perp}^{\text{kin}}\rangle \ne 0$.

We would like to emphasize that the stationary Landau states are not the most general solution to the problem and there is a family of other states with different quantum numbers, including \textit{the non-stationary states} that also have $\langle \hat{L}_z\rangle = \ell$ \cite{Bagrov2002}. Nevertheless, in order to fully describe quantum dynamics of all the vortex electron's observables in a magnetic lens it is enough to know the incoming free states, the stationary states, and to apply the boundary conditions together with the quantum equations of motion.

\subsection{In axially symmetric crossed fields}\label{Symm}

The conservation of the canonical angular momentum $\ell$ takes place for more general axially symmetric and not necessarily homogeneous fields such as a combination of magnetic and electric lenses\cite{Reiser}.
Indeed, the condition $[\hat{H}, \hat{L}_z^{\text{can}}] = 0$ holds when neither ${\bm A}$ nor $A^0$ depends on $\phi$.
The simplest example is when along with the solenoid field (\ref{Adef}) we also have a constant and homogeneous longitudinal electric field ${\bm E} = \{0, 0, E\}$.
Clearly, the rms-radius $\langle \rho^2 \rangle(t)$ from Eq.(\ref{r0gen}) is not changed, so such a field can accelerate vortex particles. 
This rms-radius \textit{is} changed, however, when the electric field has a radial component $E_{\rho}$ (electric lens).

Let us suppose that along with the longitudinal magnetic field and the longitudinal (accelerating) electric field we have an \textit{inhomogeneous} radial component of the electric field, 
\begin{eqnarray}
& \displaystyle
{\bm E} = E_{\rho}(\rho)\,{\bm e}_{\rho} + E_{z}\,{\bm e}_{z},\ {\bm H} = \{0, 0, H\}.
\label{lE}
\end{eqnarray}
where ${\bm e}_{\rho} = {\bm \rho}/\rho,\, {\bm e}_{z} = \{0,0,1\}$. In \textit{a linear approximation}, we suppose that the radial field is absent on the axis, $E_{\rho}(0) = 0$, and so (see Sec.3.3.2 in Ref.\cite{Reiser})
\begin{eqnarray}
& \displaystyle
E_{\rho}(\rho) \approx \rho E_{\rho}^{\prime}
\label{Erholin}
\end{eqnarray}
where $E_{\rho}^{\prime} \equiv \partial E_{\rho}(0)/\partial\rho$ can be either positive or negative. In the same approximation, a radial component of the magnetic field (\ref{Hrho}) 
can also be added to (\ref{lE}) for a solenoid of a finite length, which does not however change the main conclusions below. 
An example of the linear field (\ref{lE}),(\ref{Erholin}) is an azimuthally symmetric \textit{Penning trap} (as well as its generalizations such as a Penning–Malmberg trap) with the following scalar potential 
(see, for instance, \cite{Eseev}):
\begin{eqnarray}
& \displaystyle
A^0(\rho,z) = a\, (\rho^2 - 2z^2),\ a = \text{const},\ E_{\rho}(\rho) = -2a \rho.
\label{Penning}
\end{eqnarray}

The quantum equations of motion look now as follows (cf. Eq.(\ref{Heiseq})):
\begin{eqnarray}
& \displaystyle \frac{\partial^2\langle\rho^2\rangle(t)}{\partial t^2} = 2\langle{\bm u}_{\perp}^2\rangle(t) + 2\frac{\omega_c}{m}\,\ell - \langle\rho^2\rangle(t) \left(\omega_c^2 - \frac{2eE_{\rho}^{\prime}}{m}\right),\cr
& \displaystyle \frac{\partial^2\langle{\bm u}_{\perp}^2\rangle(t)}{\partial t^2} = \frac{eE_{\rho}^{\prime}}{m}\,\frac{\partial^2\langle\rho^2\rangle(t)}{\partial t^2}.
\label{HeiseqE}
\end{eqnarray}
where there is an additional term with $E_{\rho}^{\prime}$ and $\langle{\bm u}_{\perp}^2\rangle$ is no longer constant. 
The second equation together with the condition $\langle{\bm u}_{\perp}^2\rangle(t) = \langle{\bm u}_{\perp}^2\rangle$ at $E_{\rho}=0$ yield
\begin{eqnarray}
& \displaystyle \langle{\bm u}_{\perp}^2\rangle(t) = \langle{\bm u}_{\perp}^2\rangle + \frac{eE_{\rho}^{\prime}}{m}\,\langle\rho^2\rangle(t).
\label{u2E}
\end{eqnarray}
Then the first equation looks as follows:
\begin{eqnarray}
& \displaystyle \frac{\partial^2\langle\rho^2\rangle(t)}{\partial t^2} = 2\langle{\bm u}_{\perp}^2\rangle + 2\frac{\omega_c}{m}\,\ell - \langle\rho^2\rangle(t) \left(\omega_c^2 - \frac{4eE_{\rho}^{\prime}}{m}\right) = \cr
& \displaystyle = \omega_c^2 \langle\rho^2\rangle_{\text{st}} - \langle\rho^2\rangle(t) \left (\omega_c^2 - \frac{4eE_{\rho}^{\prime}}{m}\right).
\label{HeiseqE2}
\end{eqnarray}
where $\langle\rho^2\rangle_{\text{st}}$ is from Eq.(\ref{r0}).

The solution to this equation crucially depends on the sign of $\omega_c^2 - 4eE_{\rho}^{\prime}/m$. For a simplest \textit{electrostatic lens} with $H = 0$ (say, an \textit{einzel lens}), 
the solution is oscillatory (the lens is focusing) only when 
$$
eE_{\rho}^{\prime} < 0,
$$
i.e., either $e < 0, E_{\rho}^{\prime} > 0$ or vice versa. It is
\begin{eqnarray}
& \displaystyle \langle\rho^2\rangle(t) = \frac{\varepsilon_{\perp}}{|eE_{\rho}^{\prime}|} + \alpha_1\, \cos \left(t\sqrt{\frac{4}{m}\,|eE_{\rho}^{\prime}|} + \alpha_2\right), \cr
& \displaystyle \text{or}\quad \langle\rho^2\rangle(t) = \frac{\varepsilon_{\perp}}{|eE_{\rho}^{\prime}|} + \left(\langle\rho^2\rangle^{\text{free}} - \frac{\varepsilon_{\perp}}{|eE_{\rho}^{\prime}|}\right)\, \cos \left(t\sqrt{\frac{4}{m}\,|eE_{\rho}^{\prime}|}\right)
\label{rhoE2st}
\end{eqnarray}
when a free particle enters the lens at $t=0$, see Fig.\ref{Transm}. In the stationary regime (a thick lens), we have 
\begin{eqnarray}
& \displaystyle
\overline{\langle\rho^2\rangle(t)} = \varepsilon_{\perp}/|eE_{\rho}^{\prime}|.
\label{rhomeanst}
\end{eqnarray}
In the opposite case with $eE_{\rho}^{\prime} > 0$, the rms-radius grows with time -- the lens is \textit{globally defocusing}.

When the magnetic field is not vanishing, the lens is focusing and the mean emittance is conserved only when (in accord with Eq.(25.35) in Ref.\cite{ST})
\begin{eqnarray}
& \displaystyle
\omega_c^2 - 4eE_{\rho}^{\prime}/m > 0,
\label{stable}
\end{eqnarray}
which can be realized even for $eE_{\rho}^{\prime} > 0$. The corresponding solution is
\begin{eqnarray}
& \displaystyle \langle\rho^2\rangle(t) = \frac{\omega_c^2}{\omega_c^2 - 4eE_{\rho}^{\prime}/m}\,\langle\rho^2\rangle_{\text{st}} + \alpha_1\, \cos \left(t\sqrt{\omega_c^2 - 4eE_{\rho}^{\prime}/m} + \alpha_2\right), \cr
& \displaystyle \text{or}\quad \langle\rho^2\rangle(t) = \frac{\omega_c^2}{\omega_c^2 - 4eE_{\rho}^{\prime}/m}\,\langle\rho^2\rangle_{\text{st}} + \cr
& \displaystyle + \left(\langle\rho^2\rangle^{\text{free}} - \frac{\omega_c^2}{\omega_c^2 - 4eE_{\rho}^{\prime}/m}\,\langle\rho^2\rangle_{\text{st}}\right)\, \cos \left(t\sqrt{\omega_c^2 - 4eE_{\rho}^{\prime}/m}\right).
\label{rhoHE2st}
\end{eqnarray}
when a particle enters the field at $t=0$. For a thin lens, the time interval during which the particle stays in the field is shorter than the oscillation period
\begin{eqnarray}
& \displaystyle T_{HE} = \frac{2\pi}{\sqrt{\omega_c^2 - 4eE_{\rho}^{\prime}/m}},
\label{TEH}
\end{eqnarray}
and we also have an effective diffraction time $t_d$, a $\hat{\beta}$-function, and a Gouy phase $\varphi^G = \pm \arctan t/t_d = \pm \int dt/\hat{\beta}(t) \approx \pm t/t_d$,
\begin{eqnarray}
& \displaystyle \langle\rho^2\rangle(t) = \langle\rho^2\rangle^{\text{free}} \left (1 \pm \frac{t^2}{t_d^2}\right),\quad t_d = \frac{T_{HE}}{\pi \sqrt{2}}\frac{\sqrt{\langle\rho^2\rangle^{\text{free}}}}{\sqrt{\left|\langle\rho^2\rangle^{\text{free}} - \frac{\omega_c^2}{\omega_c^2 - 4eE_{\rho}^{\prime}/m}\,\langle\rho^2\rangle_{\text{st}}\right|}}.
\label{TdEH}
\end{eqnarray}
Importantly, at $\omega_c^2 - 4eE_{\rho}^{\prime}/m \to +0$ the oscillations get ``frozen'', $T_{HE} \to \infty$, and the Gouy phase becomes positive, which means that the packet spreads and the lens is defocusing not only for $\omega_c^2 - 4eE_{\rho}^{\prime}/m < 0$.

\begin{figure}
\centering
    \includegraphics[width=10cm]{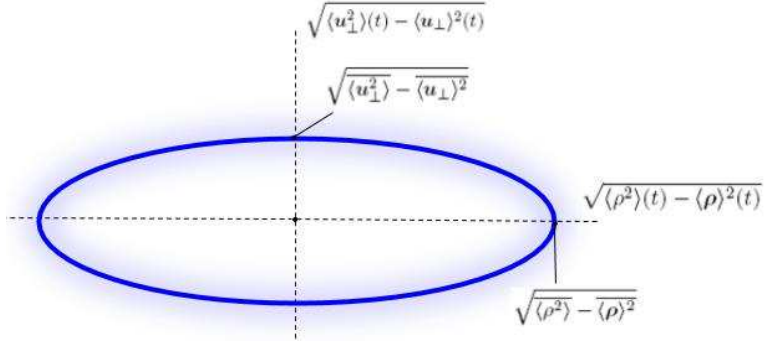}
    \center
		\caption{In focusing, axially symmetric fields, the transverse phase space of the wave packet is blurred due to quantum broadening of the classical trajectories 
		and the mean slope angle is vanishing. The mean area of the ellipse defines the conserved emittance of the particle.}
\label{Ellipse_Field}
\end{figure}

To estimate the emittance and the quality factor, we also need the solutions of the equations of motion
\begin{eqnarray}
& \displaystyle \frac{\partial^2\langle{\bm \rho}\rangle}{\partial t^2} = \frac{e}{m}\, \langle {\bm E}_{\rho} \rangle + \frac{e}{m}\, \langle {\bm u}\times{\bm H}\rangle,\ \langle {\bm E}_{\rho} \rangle = E_{\rho}^{\prime}\,\langle{\bm \rho}\rangle.
\label{eqmHE}
\end{eqnarray}
For the focusing lens, they are
\begin{eqnarray}
& \displaystyle \langle x\rangle(t) = \delta_1\,\cos\left(\frac{t}{2}\,\left(-\omega_c + \sqrt{\omega_c^2 - 4eE_{\rho}^{\prime}/m}\right) + \varphi_1\right) + \cr
& \displaystyle + \delta_2\,\cos\left(\frac{t}{2}\,\left(-\omega_c - \sqrt{\omega_c^2 - 4eE_{\rho}^{\prime}/m}\right) + \varphi_2\right),\cr
& \displaystyle \langle y\rangle(t) = \delta_1\,\sin\left(\frac{t}{2}\,\left(-\omega_c + \sqrt{\omega_c^2 - 4eE_{\rho}^{\prime}/m}\right) + \varphi_1\right) + \cr
& \displaystyle + \delta_2\,\sin\left(\frac{t}{2}\,\left(-\omega_c - \sqrt{\omega_c^2 - 4eE_{\rho}^{\prime}/m}\right) + \varphi_2\right),
\label{eqmHESol}
\end{eqnarray}
where $\delta_1,\delta_2,\varphi_1,\varphi_2$ are defined by the initial conditions. The matrix of central moments
\begin{eqnarray}
& \displaystyle Q_{ij}(t) = 
\begin{pmatrix}
\langle\rho^2\rangle - \langle{\bm \rho}\rangle^2 & \langle{\bm \rho}\cdot{\bm u}_{\perp}\rangle - \langle{\bm \rho}\rangle\cdot\langle{\bm u}_{\perp}\rangle\\
\langle{\bm \rho}\cdot{\bm u}_{\perp}\rangle - \langle{\bm \rho}\rangle\cdot\langle{\bm u}_{\perp}\rangle & \langle{\bm u}_{\perp}^2\rangle - \langle{\bm u}_{\perp}\rangle^2
\end{pmatrix}
\label{QEH}
\end{eqnarray}
becomes diagonal after averaging over the oscillation period,
\begin{eqnarray}
& \displaystyle \overline{Q_{ij}(t)} = 
\begin{pmatrix}
\overline{\langle\rho^2\rangle} - \overline{\langle{\bm \rho}\rangle^2} & 0 \\
0 & \overline{\langle{\bm u}_{\perp}^2\rangle} - \overline{\langle{\bm u}_{\perp}\rangle^2}
\end{pmatrix}
\label{QEHav}
\end{eqnarray}
where
\begin{eqnarray}
& \displaystyle \overline{\langle\rho^2\rangle(t)} = \frac{\omega_c^2}{\omega_c^2 - 4eE_{\rho}^{\prime}/m}\,\langle\rho^2\rangle_{\text{st}},\ \overline{\langle{\bm \rho}\rangle^2} = \delta_1^2 + \delta_2^2,\cr
& \displaystyle \overline{\langle{\bm u}_{\perp}^2\rangle(t)} = \langle{\bm u}_{\perp}^2\rangle + \frac{eE_{\rho}^{\prime}}{m}\,\frac{\omega_c^2}{\omega_c^2 - 4eE_{\rho}^{\prime}/m}\,\langle\rho^2\rangle_{\text{st}},\cr
& \displaystyle \overline{\langle{\bm u}_{\perp}\rangle^2} = \frac{1}{4}\left(\delta_1^2\, \left(\omega_c - \sqrt{\omega_c^2 - 4eE_{\rho}^{\prime}/m}\right)^2 + \delta_2^2\, \left(\omega_c + \sqrt{\omega_c^2 - 4eE_{\rho}^{\prime}/m}\right)^2\right).
\label{rumean}
\end{eqnarray}
The effective emittance $\epsilon_{\rho} = \sqrt{\det \overline{Q_{ij}(t)}}$ and the quality factor $M = \epsilon_{\rho}/\lambda_c$ are conserved (see Fig.\ref{Ellipse_Field}), 
they grow with $\ell$ together with $\langle\rho^2\rangle_{\text{st}}$, while the latter radius is to be calculated by using the quantum states in the field (\ref{lE}) given in Ref.\cite{ST}.
Remarkably, the use of the more intricate non-stationary states \cite{Bagrov2002} is not needed to fully describe quantum dynamics in the focusing electric and magnetic lenses. 

Thus twisted wave packets can be safely transmitted through a combination of axially symmetric and linear magnetic and electric lenses without changing their OAM, see Fig.\ref{Transm}. 
The rms radius and the rms velocity \textit{can} be changed by the lenses and their final values depend on the moment of time, i.e. on the phase, at which the particle leaves the lens.

\subsection{Born in a solenoid and the quantum Busch theorem}\label{BuschEl}

\begin{figure}[ttt]
\centering
    \includegraphics[width=9cm]{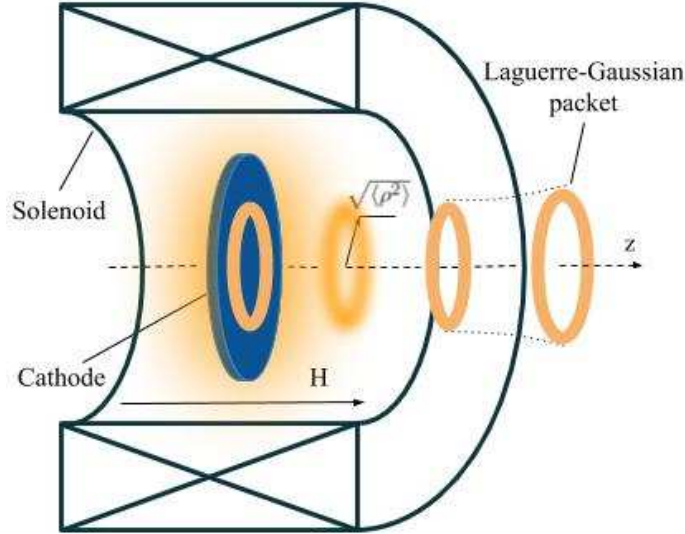}
    \center
		\caption{Vortex electrons born from a magnetized cathode. The OAM $\ell$ is due to the quantized flux of the magnetic field through an area $\pi \langle \rho^2\rangle$ of the wave packet according to the quantum Busch theorem (\ref{Busch}). The electrons can further be accelerated -- for instance, by the longitudinal electric field -- and when leaving the solenoid the electron's quantum state turns into the free Laguerre-Gaussian packet (\ref{LGSt}).}
\label{Solenoid}
\end{figure}

Let us suppose that at $t = 0, \langle z\rangle = 0$ a non-rotating wave-packet with 
\begin{eqnarray}
& \displaystyle
\langle \hat{L}_z^{\text{kin}}\rangle = 0
\label{Lzvanish}
\end{eqnarray} 
is generated inside the solenoid -- say, emitted from a thermocathode or a photocathode. As can be seen from Eq.(\ref{Lzkin}), the canonical angular momentum $\ell$ is necessarily \textit{not vanishing} 
and it has the same sign as $e H$ ($H$ can be negative in this section),
\begin{eqnarray}
& \displaystyle \ell= \frac{e H}{2}\,\langle \rho^2\rangle = 2\,\sgn(e)\, \frac{\langle \rho^2\rangle}{\rho_H^2}  = \frac{1}{2}\,\sgn(e)\,\frac{\langle \rho^2\rangle}{\lambda_c^2}\,\frac{H}{H_c},
\label{Busch}
\end{eqnarray}
where $\sqrt{\langle \rho^2\rangle} \equiv \sqrt{\langle \rho^2\rangle(0)}$ is the packet's rms-radius in the region of generation -- say, on a surface of the cathode, see Fig.\ref{Solenoid}. 
The flux of the magnetic field $\langle \Phi\rangle = H \pi \langle \rho^2\rangle$ through an area $\pi \langle\rho^2\rangle$ of the wave packet at a point of generation is quantized, analogously to Eq.(\ref{Phimean}).

An instantaneous emission of an electron implies that the duration of the process $\delta t \sim 1/\delta\varepsilon$ is much shorter than the cyclotron period $T_c$. 
So the energy uncertainty must be large (see Eq.(\ref{TcHc})),
\begin{eqnarray}
& \displaystyle 
\delta \varepsilon \gg m\, \frac{H}{H_c} \sim (10^{-4} - 10^{-2})\, \text{eV}
\label{Ineqeps}
\end{eqnarray}
for $H \sim 0.1 - 10$ T. The typical values of $\delta \varepsilon \sim 0.1 - 0.5$ eV for field emission and photoemission \cite{Ehberger} satisfy this condition.

In accelerator physics, Eq.(\ref{Busch}) is called \textit{the Busch theorem} \cite{Busch, Reiser, Groening2018} and it relates the field strength $H$ on a magnetized cathode, the beam rms-radius at the point of generation, and the resultant non-integer AM. Here, we deal with a single-particle wave packet and $\langle...\rangle$ is a quantum averaging over the Bohmian trajectories, and Eq.(\ref{Busch}) can be called \textit{the quantum Busch theorem}. The presence of an electric field \textit{does not influence} this theorem and, therefore, if a vortex packet with the canonical AM $\ell$ is born inside a solenoid it can further be accelerated by the field $E_z$ to an anode and its emittance stays nearly the same. When leaving the field -- say, through a round aperture in the anode -- the particle stays twisted due to the conservation of the canonical AM.

When emitted from a thermocathode or a photocathode, the electrons' transverse coherence length \textit{at room temperature} is
$$
\sqrt{\langle \rho^2\rangle} \sim 1\,\text{nm}
$$
for tungsten \cite{Cho, Ehberger, Lat}, so the minimal field strength to get an electron with $\ell = 1$ is 
\begin{eqnarray}
& \displaystyle 
H \sim 10^2\,\text{T},
\label{Hmin}
\end{eqnarray}
while it is $H \sim 1$\,T for $\sqrt{\langle \rho^2\rangle} \sim 10$\,nm. The transverse coherence of the order of $10-100\,\text{nm}$ can be achieved by using cryogenically cooled field emitters, photocathodes or cold-atom electron sources \cite{Cho, Cho2013, Ehberger, McCulloch}. The cold-atom electron sources reveal the Maxwellian momentum distribution of the emitted electrons \cite{McCulloch} with
\begin{eqnarray}
& \displaystyle 
\sqrt{\langle \rho^2\rangle} = \frac{1}{\sqrt{m\, k_B T}},
\label{StMaxw}
\end{eqnarray}
where $k_B$ is the Bolzmann constant. This yields some $1.7$ nm at room temperature, in accord with the results for field emitters and photocathodes \cite{Cho, Ehberger, Lat}.
For lower temperatures, however, the temperature dependence of the transverse coherence length for field emission reveals an approximately inverse $\sqrt{\langle \rho^2\rangle} \propto 1/T$ dependence 
rather than the Maxwellian $\sqrt{\langle \rho^2\rangle} \propto 1/\sqrt{T}$ one. Indeed, decrease of the temperature from the room value to $78$ K resulted in a corresponding increase 
of the electron coherence in Ref.\cite{Cho},
\begin{eqnarray}
& \displaystyle 
\sqrt{\langle \rho^2\rangle}_{T=78\,\text{K}}/\sqrt{\langle \rho^2\rangle}_{T\approx 295\,\text{K}} = 295/78 \approx 3.8.
\label{StFD}
\end{eqnarray}
This hints that the Fermi-Dirac distribution should be used for such partially coherent sources as field emitters and photocathodes.

A minimal inelastic mean free path of electrons in a 2D electron gas in metals varies from $10$ nm to some $100$ nm in the temperature range of $3.5 - 178$ K,
and it reveals an inverse $1/m^*k_B T$ dependence due to the Fermi-Dirac statistics with an effective mass $m^*$\cite{Jean}. 
One can expect, therefore, that the similar $1/T$ dependence (\ref{StFD}) can take place for the coherence length of electrons emitted from the surface of metals, which would yield 
$$
\sqrt{\langle \rho^2\rangle} \sim 30\,\text{nm} 
$$ 
at $T = 10$ K. In this case, the field of $H\sim 0.1$ T would suffice for creation of an electron with $\ell \sim 1$.
In an optimistic scenario, an electron with $\sqrt{\langle \rho^2\rangle} \sim 100$\,nm and the kinetic energy of several eV ($u\sim 10^{-3}\,c$) is coherently emitted from a cooled source. 
Its emission time is roughly $100\,\text{nm}/u \sim 10^{-13}$\,sec., which is still smaller than $T_c$ from (\ref{TcHcqt}), and the process can be treated as instantaneous. 
The inverse of this time defines the minimum energy width $\delta \varepsilon \lesssim 10^{-2}$\,eV and so $\delta\varepsilon/\varepsilon < 10^{-3}$ at the emission region.  

The requirement to have a cathode at low temperature can be circumvented by using a photocathode illuminated by \textit{a Laguerre-Gaussian beam} of optical twisted photons as we discuss in Ref.\cite{PRA2020}.
If a detector does not differentiate the azimuthal angles at which an electron is emitted, the electrons will be coherent over all azimuthal angles 
and the electron coherence length will match that of the incident photon beam, which can easily be much larger than $1\,\mu\text{m}$. Thus the generation of vortex electrons with $\ell \gg 1$ 
and the subsequent acceleration of them to relativistic energies seem feasible in this scheme.

\subsection{Production of vortex ions, protons, and other particles}\label{BuschI}

The quantum Busch theorem implies that the magnetized cathode technique can be adapted to generate other types of charged particles with OAM, 
such as protons and antiprotons, light and heavy ions, positrons, muons, etc. For a particle with a charge $q = Ze$, the Busch theorem is obtained from Eq.(\ref{Busch}) by $e \to Ze$ 
and it can be presented as follows:
\begin{eqnarray}
|\ell| \approx 1.5 \times 10^{-3}\,|Z|\,\langle \rho^2\rangle [\text{nm}^2]\, |H| [\text{T}],
\label{Eqell}
\end{eqnarray}
where $\sqrt{\langle \rho^2\rangle}$ and $H$ are measured in nanometers and in Tesla, respectively. So the OAM grows with $|Z|$. 
If the transverse coherence length of an electron emitted from a cathode at a certain temperature scales as
\begin{eqnarray}
& \displaystyle 
\sqrt{\langle \rho^2\rangle} \propto 1/mT,
\label{rhosc}
\end{eqnarray}
then for a source of fermions \textit{of a different mass} at the same temperature we have
\begin{eqnarray}
& \displaystyle 
T_1=T_2:\ \frac{\sqrt{\langle \rho^2\rangle}(m_1)}{\sqrt{\langle \rho^2\rangle}(m_2)} = \frac{m_2}{m_1}.
\label{rhoscratio}
\end{eqnarray}
In particular, if $m_2 = m_p$ is a proton mass then the proton wave packet is \textit{some $1840$ times narrower} than that of electron.
At room temperature, this yields 
\begin{eqnarray}
& \displaystyle 
\sqrt{\langle \rho^2\rangle}(m_p) \lesssim 1\, \text{pm}.
\label{rhoscratio2}
\end{eqnarray}
The scaling (\ref{rhoscratio}) and the estimate (\ref{rhoscratio2}) coincide with those obtained for an ultrarelativistic bunch of non-interacting fermions. Indeed, Eq.(61) in Ref.\cite{PRD} yields
\begin{eqnarray}
& \displaystyle 
\frac{\delta\varepsilon_1}{\varepsilon_1}=\frac{\delta\varepsilon_2}{\varepsilon_2}:\ \frac{\sqrt{\langle \rho^2\rangle}(m_1)}{\sqrt{\langle \rho^2\rangle}(m_2)} \geq \frac{\lambda_c(m_1)}{\lambda_c(m_2)} = \frac{m_2}{m_1},
\label{rhoscratio3}
\end{eqnarray}
and so 
\begin{eqnarray}
& \displaystyle 
\sqrt{\langle \rho^2\rangle}(m_p) \geq \frac{\lambda_c(m_p)}{\delta\varepsilon/\varepsilon} \sim 0.2\, \text{pm}\ \text{for}\ \delta\varepsilon/\varepsilon \sim 10^{-3},
\label{rhoscratio4}
\end{eqnarray}
i.e. the packet is paraxial, $\sqrt{\langle \rho^2\rangle}(m_p) \gg \lambda_c(m_p) = \hbar/m_p c$. The lower temperatures correspond to lower values of $\delta\varepsilon/\varepsilon$ and hence to the larger coherences.

\begin{figure}[ttt]
\centering
    \includegraphics[width=11cm]{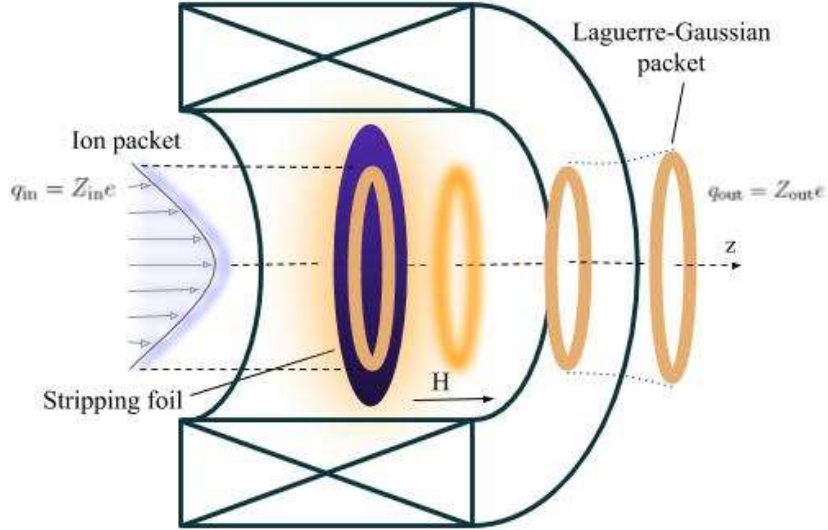}
    \center
		\caption{Vortex ions or protons born from a magnetized stripping foil with the OAM defined by the ion version of the Busch theorem, Eq.(\ref{Buschfoil}).
		The foil can be made in a form of a ring.}
\label{Foil}
\end{figure}

One can employ \textit{a stripping foil} immersed in the solenoid field at some point $\langle z\rangle$, applied when changing the charge state of ion beams \cite{Groening2014, Groening2017, Hwang2014, Groening2018}. It can be used to produce both the vortex ions -- from H$^-$ to uranium -- and the vortex protons. Neglecting multiple scattering in the foil, the second beam moments are continuous together with the emittance, which is often the case for low-intensity ion beams \cite{Groening2017, Groening2018}, so the rms width $\sqrt{\langle \rho^2\rangle}$ of the outgoing ion packet is defined by that of the incoming ion.
The canonical angular momentum experiences a jump from $0$ to $\ell$ at the foil due to the change of charge from $q_{\text{in}} = Z_{\text{in}} e$ to $q_{\text{out}} = Z_{\text{out}} e$, 
while the kinetic AM stays the same, see Fig.\ref{Foil}. 
So the quantum Busch theorem in this case is
\begin{eqnarray}
& \displaystyle \ell = \frac{q_{\text{out}} - q_{\text{in}}}{2}\, H\, \langle \rho^2\rangle = \left(Z_{\text{out}} - Z_{\text{in}}\right)\frac{eH}{2}\, \langle\rho^2\rangle.
\label{Buschfoil}
\end{eqnarray}
where $\langle \rho^2\rangle$ is taken at the point $\langle z\rangle$ where the foil is installed. Compared to the Busch theorem for electrons (\ref{Busch}), 
we have an enhancement due to the factor $Z_{\text{out}} - Z_{\text{in}}$ but the transverse coherences seem to be smaller than those for electrons.

The non-relativistic packets rapidly spread while propagating, 
which is why $\sqrt{\langle \rho^2\rangle}\left(\langle z\rangle\right)$ at the foil can be much larger than $\sqrt{\langle \rho^2\rangle}(0)$ at an ion source 
if the distance $\langle z\rangle$ between the source and the foil is much larger than the Rayleigh length (see Eq.(\ref{tdem}))
\begin{eqnarray}
& \displaystyle z_R = \beta \frac{\langle \rho^2\rangle(0)}{M \lambda_c},\, \lambda_c = \hbar/m_i c, 
\label{Ralion}
\end{eqnarray}
where $m_i$ is the ion mass, $M$ is the packet's quality factor. For electrons with $\sqrt{\langle \rho^2\rangle} \sim 1$ nm, the Rayleigh length is 
\begin{eqnarray}
& \displaystyle z_R^{(e)} \approx 2.5\, \beta\, [\mu\text{m}], 
\label{Raylel}
\end{eqnarray}
whereas for protons with $\sqrt{\langle \rho^2\rangle} \sim 1$ pm we have  
\begin{eqnarray}
& \displaystyle z_R^{(p)} \approx 5\, \beta\, [\text{nm}],
\label{RaylP}
\end{eqnarray}
and yet smaller values for heavy ions as 
\begin{eqnarray}
& \displaystyle
z_R \propto \langle \rho^2\rangle/\lambda_c \propto 1/m_i
\label{zRmest}
\end{eqnarray} 
if the scaling (\ref{rhosc}) holds valid. Note that the Maxwellian scaling (\ref{StMaxw}) predicts that $z_R$ does not depend on the mass at all, which is definitely wrong. 
Clearly, for any macroscopic distance -- say, of several mm and larger -- the incident packet with $\beta \sim 10^{-3}-10^{-1}$ can be considered to be spatially coherent 
\textit{over the entire width} of the foil (see Fig.\ref{Foil}), provided that no accelerating or focusing fields are applied between the source and the foil 
with a possible exception of a collimator. The use of a collimator could guarantee that the beam intensity is low and no separation of the final ions of different charges is required. 
Indeed, as seen from the previous sections the quantum packet \textit{does not spread} on average in axially symmetric electric and magnetic fields,
which is why any accelerating section between the source and the foil would leave the packet's rms radius nearly the same.
To avoid this, one could install a short \textit{drift section} with no fields whatsoever after the collimator and the accelerating section to let the accelerated ion packet freely propagate and spread
before hitting the foil. 

Thus twisting heavy charged particles with a magnetized stripping foil can be \textit{even more effective} than it is for electrons with a magnetized cathode.
To produce individual packets of vortex particles, the beam intensity and the current must be low enough to exclude the space charge effects.
One can also envisage a foil made in a form of \textit{a spiral or a ring}, which would facilitate the production of vortex ions. 
Alternatively, gas and liquid type strippers can also be employed -- see, for instance, Ref.\cite{Hwang2014}.

Let us briefly discuss how the scattering in the foil could alter these predictions. The spreading packet with the OAM $\ell$ can be characterized with a following opening angle:
\begin{eqnarray}
\tan\alpha \approx \alpha = \frac{\sqrt{\langle\rho^2\rangle}(z_R)-\sqrt{\langle\rho^2\rangle}}{z_R} \approx 0.41 \frac{\sqrt{\langle\rho^2\rangle}}{z_R} \sim \frac{|\ell|}{\beta}\, 10^{-4},
\label{alpha}
\end{eqnarray}
for $\lambda_c/\sqrt{\langle\rho^2\rangle} \sim 10^{-4}$. If $\beta < 10^{-1}$ this angle is larger than the angular straggling in the foil of $0.19$ mrad 
reported in Ref.\cite{Hwang2014}. The angular straggling of $\delta\alpha \sim 1$ mrad results in the variation of the OAM $\delta |\ell| \sim 1$  for $\beta \sim 0.1$.
So scattering in a foil, gas, or liquid stripper leads to broadening of the OAM spectrum, which can be safely neglected for $|\ell| \gg 1$,
provided that the beam quality does not degrade much and the space-charge effects can be neglected.

Other possibilities to prepare and store vortex ions, positrons, antiprotons, and other exotic particles include the use of \textit{a Penning trap} and its modifications such as a Penning–Malmberg trap because the canonical AM is also an integral of motion for them (see Sec.\ref{Symm}). One can employ a cathode with a large spatial coherence of the emitted electrons 
exposed to the magnetic field stronger than $1$ T like in \textit{a Penning negative ion source} or in \textit{a Penning ion gauge source} used, for instance, at GSI in Darmstadt \cite{GSI}.
A magnetized photocathode with an incident beam of twisted photons can also be used. 

If a vortex particle is generated inside the trap, its OAM would hold provided that collisions 
in the plasma can be neglected. So the number of twisted particles stored in a trap cannot be large due to quantum decoherence. Another problem is how to extract such particles from the trap. 
For negative ions, the use of a caesium-coated \textit{ring-shaped cathode} can be envisaged. The ions can be extracted through a round aperture made in the center of the ring 
and so the vortex ions, which have larger spatial coherence, can be separated from the untwisted ions by varying the radius of the slit and letting the untwisted particles leave the trap.

Finally, note that the Busch theorem (\ref{Eqell}) does not specify the physical process that led to the creation of a particle with the canonical AM $\ell$.
Therefore it holds valid for other processes in magnetic fields, for instance, for ionization of cold atoms in magneto-optical traps. One can photoionize cold Rydberg atoms with the radii up to 1 $\mu\text{m}$
in a magnetic field and the resulting electron and the Rydberg ion will acquire a canonical angular momentum. However a dedicated study of the spatial coherence of photoelectrons emitted from such atoms is needed as the results of Ref.\cite{Anderson} hint that the electron coherence length can be but a few nanometers.

\section{Conclusion}\label{Concl}

Due to the finite coherence length, quantum wave packets behave in external fields very similar to beams of classical non-interacting particles in the spirit of Bohmian mechanics.
In a sense, a wave packet is analogous to a beam of point particles with the total delocalized charge $e$ or $Ze$ spread over the entire coherence length.
That is why quantum dynamics of the wave packets in electric and magnetic lenses is very similar, although not identical, to classical dynamics of beams,
described by the Courant-Snyder formalism. In particular, the intrinsic orbital angular momentum of vortex particles is also conserved in axially symmetric and linear lenses and in Penning traps, 
exactly like the angular momentum of classical rotating beams. This analogy is not complete as the packet's emittance oscillates in the focusing lenses around a mean value 
defined by the stationary quantum states in the field and the OAM that a particle acquires in a magnetized cathode or in a stripping medium is quantized due to the quantum Busch theorem.

Thus the techniques of particle optics in linear accelerators and Penning traps can be effectivly used to accelerate, focuse, trap, and store the vortex particles.
The non-symmetric quadrupole lenses, in their turn, can be used to steer and focuse particles without an OAM (for instance, the Hermite-Gaussian packets \cite{Scht}),
while focusing of a vortex packet in a short quadrupole lens would result in a broadening of the OAM spectrum keeping the central value.
The methods to generate classical beams of ions and protons with angular momentum can also be applied for the production of quantum vortex states of particles heavier than electron.
The development of these methods could significantly widen the range of applications of the twisted wave packets, especially in hadronic and heavy ion physics.

I am grateful to V.\,G.~Bagrov, P.\,Kazinski, A.~Pupasov-Maksimov, V.\,G.~Serbo, and, especially, to K.~Floettmann for fruitful discussions and criticism. 
This work is supported by the Russian Science Foundation (Project No.\,17-72-20013).


\begin{thebibliography}{80}

\bibitem{Bliokh}
K.Y.~Bliokh, I.P.~Ivanov, G.~Guzzinati, L.~Clark, R.~Van Boxem, A.~B\'ech\'ed, R.~Juchtmans, M.A.~Alonso, P.~Schattschneider, F.~Nori, J.~Verbeeck,
Theory and applications of free-electron vortex states, Phys. Rep. {\bf 690}, 1 (2017).

\bibitem{Lloyd}
S.\,M.~Lloyd, M.~Babiker, G.~Thirunavukkarasu, and J.~Yuan, Electron vortices: Beams with orbital angular momentum, Rev. Mod. Phys. 89, 035004 (2017).

\bibitem{Ivanov11}
I.\,P.~Ivanov, Colliding particles carrying non-zero orbital angular momentum, Phys. Rev. D {\bf 83}, 093001 (2011).

\bibitem{Ivanov12}
I.\,P.~Ivanov, Measuring the phase of the scattering amplitude with vortex beams, Phys. Rev. D {\bf 85}, 076001 (2012).

\bibitem{PRA12}
D.\,V.~Karlovets, Electron with orbital angular momentum in a strong laser wave, Phys. Rev. A 86, 062102 (2012).

\bibitem{Bliokh12}
K.\,Y.~Bliokh, P.~Schattschneider, J.\,Verbeeck, and F.~Nori, Electron Vortex Beams in a Magnetic Field: A New Twist on Landau Levels
and Aharonov-Bohm States, Phys. Rev. X 2, 041011 (2012).

\bibitem{McMorran}
G.\,M.~Gallatin, B.~McMorran, Propagation of vortex electron wave functions in a magnetic field, Phys. Rev. A 86, 012701 (2012).

\bibitem{Beche2014} 
B\'ech\'e, R.~van~Boxem, G.~van~Tendeloo, J.~Verbeeck, Magnetic monopole field exposed by electrons, Nat. Phys. \textbf{10}, 26 (2014).

\bibitem{FA1}
C.R. Greenshields, R.L. Stamps, S. Franke-Arnold, S.M. Barnett, Is the angular momentum of an electron conserved in a uniform magnetic field? 
Phys. Rev. Lett. 113 (2014) 240404.

\bibitem{FA2}
C. Greenshields, S. Franke-Arnold, R.L. Stamps, Parallel axis theorem for free-space electron wavefunctions, New J. Phys. 17 (2015) 093015.

\bibitem{Serbo15}
V.\,G.~Serbo, I.~Ivanov, S.~Fritzsche, D.~Seipt, A.~Surzhykov, Scattering of twisted relativistic electrons by atoms, Phys. Rev. A {\bf 92} 012705, (2015).

\bibitem{JHEP}
D.\,V.~Karlovets, Scattering of wave packets with phases, J. High Energy Phys. {\bf 03}, 049 (2017).

\bibitem{Ivanov16}
I.\,P.\, Ivanov, D.\, Seipt, A.\, Surzhykov, S.\, Fritzsche, Elastic scattering of vortex electrons provides direct access to the Coulomb phase, Phys. Rev. D \textbf{94}, 076001 (2016).

\bibitem{Sherwin1}
J.\,A.~Sherwin, Compton scattering of Bessel light with large recoil parameter, Phys. Rev. A \textbf{96}, 062120 (2017).

\bibitem{Sherwin2}
J.\,A.~Sherwin, Two-photon annihilation of twisted positrons, Phys. Rev. A \textbf{98}, 042108 (2018).

\bibitem{PRA18}
D. Karlovets, Relativistic vortex electrons: Paraxial versus nonparaxial regimes, Phys. Rev. A 98, 012137 (2018).

\bibitem{Moments}
D.~Karlovets, A.~Zhevlakov, Intrinsic multipole moments of non-Gaussian wave packets, 
Phys. Rev. A {\bf 99}, 022103 (2019).

\bibitem{PRA19}
D. Karlovets, Dynamical enhancement of nonparaxial effects in the electromagnetic field of a vortex electron, Phys. Rev. A 99, 043824 (2019).

\bibitem{PRD}
D.\,V.~Karlovets, V.\,G.~Serbo, Effects of the transverse coherence length in relativistic collisions, Phys. Rev. D \textbf{101}, 076009 (2020).

\bibitem{SilenkoFields}
A.\,J.~Silenko, P.~Zhang, L.~Zou, Relativistic Quantum Dynamics of Twisted Electron Beams in Arbitrary Electric and Magnetic Fields, Phys. Rev. Lett. {\bf 121}, 043202 (2018).

\bibitem{SilenkoQ}
A.\,J.~Silenko, P.~Zhang, L.~Zou, Electric quadrupole moment and the tensor magnetic polarizability of twisted electrons and a potential for their measurements, 
Phys. Rev. Lett. {\bf 122}, 063201 (2019).

\bibitem{Silenko2020}
L.~Zou, P.~Zhang, A.\,J.~Silenko, Paraxial wave function and Gouy phase for a relativistic electron in a uniform magnetic field, J. Phys. G: Nucl. Part. Phys. {\bf 47}, 055003 (2020).

\bibitem{Ivanov20201}
I.\,P.~Ivanov, N.~Korchagin, A.~Pimikov, and P.~Zhang, Doing spin physics with unpolarized particles, Phys. Rev. Lett. 124, 192001 (2020).

\bibitem{Ivanov20202}
I.\,P.~Ivanov, N.~Korchagin, A.~Pimikov, and P.~Zhang, Twisted particle collisions: a new tool for spin physics, Phys. Rev. D 101, 096010 (2020).

\bibitem{Ivanov20203}
I.\,P.~Ivanov, N.~Korchagin, A.~Pimikov, and P.~Zhang, Kinematic surprises in twisted-particle collisions, Phys. Rev. D {\bf 101}, 016007 (2020).

\bibitem{Uchida}
M.~Uchida and A.~Tonomura, Generation of electron beams carrying orbital angular momentum, Nature {\bf 464}, 737 (2010).

\bibitem{Verbeeck}
J.~Verbeeck, H.~Tian, P.~Schlattschneider, Production and application of electron vortex beams, Nature {\bf 467}, 301 (2010).

\bibitem{McMorran2}
B.~J.~McMorran A.~Agrawal, I.\,M.~Anderson, et al., Electron vortex beams with high quanta of orbital angular momentum, Science \textbf{331}, 192 (2011).

\bibitem{Reiser}
M.~Reiser, Theory and design of charged particle beams, WILEY-VCH Verlag GmbH $\&$ Co. KGaA, Weinheim (2008).

\bibitem{Burov2000}	
A.~Burov, S.~Nagaitsev, A.~Shemyakin, Y.~Derbenev, Optical principles of beam transport for relativistic electron cooling, Phys. Rev. ST Accel. Beams \textbf{3}, 094002 (2000).

\bibitem{Burov2002}	
A.~Burov, S.~Nagaitsev, Y.~Derbenev, Circular modes, beam adapters, and their applications in beam optics, Phys. Rev. E \textbf{66}, 016503 (2002).

\bibitem{Kim2003}	
K.-J.~Kim, Round-to-flat transformation of angular-momentum-dominated beams, Phys. Rev. ST Accel. Beams \textbf{6}, 104002 (2003).

\bibitem{Sun2004}	
Y.-E Sun, P.~Piot, K.-J.~Kim, N.~Barov, S.~Lidia, J.~Santucci, R.~Tikhoplav, J.~Wennerberg, Generation of angular-momentum-dominated electron beams from a photoinjector, Phys. Rev. ST Accel. Beams \textbf{7}, 123501 (2004).

\bibitem{Groening2014}
L. Groening, M. Maier, C. Xiao, L. Dahl, P. Gerhard, O.K. Kester, S. Mickat, H. Vormann, and M. Vossberg, Experimental Proof of Adjustable Single-Knob Ion Beam Emittance Partitioning, 
Phys. Rev. Lett. 113, 264802 (2014).

\bibitem{Groening2017} S.~Appel, L.~Groening, Y.~El\,Hayek, M.~Maier, C.~Xiao, Injection optimization through generation of flat ion beams, Nucl. Instrum. Meth. Phys. Res. A \textbf{866}, 36 (2017).

\bibitem{Hwang2014}
J.-G.\,Hwang, E.-S.\,Kim, H.-J.\,Kim, D.-O Jeon, Minimization of the emittance growth of multi-charge particle beams in the charge stripping section of RAON, 
Nucl. Instrum. Meth. Phys. Res. A \textbf{767}, 153 (2014).

\bibitem{Groening2018}
L. Groening, C. Xiao, M. Chung, Extension of Busch’s theorem to particle beams, Phys. Rev. Accel. Beams 21, 014201 (2018).

\bibitem{Busch} 
H.~Busch, Berechnung der Bahn von Kathodenstrahlen im axialsymmetrischen elektromagnetischen Felde, Ann. Phys. \textbf{386}, 974 (1926). 

\bibitem{Scht}
P.~Schattschneider, M.~St\"oger-Pollach, J.~Verbeeck, Novel Vortex Generator and Mode Converter for Electron Beams, Phys. Rev. Lett. {\bf 109}, 084801 (2012).

\bibitem{Floettmann2020} 
K.~Floettmann, Equivalence of Gouy and Courant-Snyder phase, Phys. Rev. A 102, 033507 (2020).

\bibitem{PRA2020} 
K.~Floettmann, D.~Karlovets, Quantum mechanical formulation of the Busch theorem, Phys. Rev. A 102, 043517 (2020).

\bibitem{Bohm} 
D.\,Bohm, A Suggested Interpretation of the Quantum Theory in Terms of "Hidden" Variables. I, Phys. Rev. 85, 166 (1952).

\bibitem{Bohm2} 
A.\,V.~Belinsky, On David Bohm's 'pilot-wave' concept, Phys.-Usp. 62, 1268 (2019).

\bibitem{Manko98}
R.~Fedele, V.\,I.~Man’ko, Quantumlike corrections and semiclassical description of charged-particle beam transport, Phys. Rev. E {\bf 58}, 992 (1998).

\bibitem{Manko99}
R.~Fedele, V.\,I.~Man’ko, Role of semiclassical description in the quantumlike theory of light rays, Phys. Rev. E {\bf 60}, 6042 (1999).

\bibitem{Dodonov2000}
V.\,V.~Dodonov, O.\,V.~Man’ko, Universal invariants of quantum-mechanical and optical systems, J. Opt. Soc. Am. A {\bf 17}, 2403 (2000).

\bibitem{Sch}
E. Schr\"odinger, Zum Heisenbergschen Unsch\"arfeprinzip, Sitzungsberichte der Preussischen Akademie der Wissenschaften, Physikalisch-mathematische Klasse 14, 296–303 (1930).

\bibitem{Mandel}
L. Mandel and E.Wolf, Optical Coherence and Quantum Optics (Cambridge University, New York, 1995).

\bibitem{Cho}
B.~Cho, T.~Ichimura, R.~Shimizu, and C.~Oshima, Quantitative Evaluation of Spatial Coherence of the Electron Beam from Low Temperature Field Emitters, Phys. Rev. Lett. 92, 246103 (2004).

\bibitem{Cho2013}
B.~Cho and C.~Oshima, Electron Beam Coherency Determined from Interferograms of Carbon Nanotubes, Bull. Korean Chem. Soc. 34, 892 (2013).

\bibitem{Lat}
T.~Latychevskaia, Spatial coherence of electron beams from field emitters and its effect on the resolution of imaged objects, Ultramicroscopy 175, 121 (2017).

\bibitem{Ehberger}
D.~Ehberger, J.~Hammer, M.~Eisele, et al., Highly Coherent Electron Beam from a Laser-Triggered Tungsten Needle Tip, Phys. Rev. Lett. 114, 227601 (2015).

\bibitem{AB}
Y.~Aharonov, D.~Bohm, Significance of Electromagnetic Potentials in the Quantum Theory, Phys. Rev. 115, 485 (1959).

\bibitem{LL2}
L.D. Landau, E.M. Lifshitz, The Classical Theory of Fields (Oxford, Pergamon, 1975).

\bibitem{Mess}
A.~Messiah, Quantum Mechanics. Vol. 1 (North-Holland, Amsterdam; Interscience, New York, 1961).

\bibitem{Rob}
H.\,P.~Robertson, The uncertainty principle, Phys. Rev. 34, 163 (1929).

\bibitem{Feist}
A.~Feist, N.~Bach, N.\,R. da Silva, et al., Ultrafast transmission electron microscopy using a laser-driven field emitter: Femtosecond resolution with a high coherence electron beam, 
Ultramicroscopy {\bf 176}, 63 (2017).

\bibitem{BLP}
V.\,B.~Berestetskii, E.\,M.~Lifshitz and L.\,P.~Pitaevskii, Quantum Electrodynamics, Oxford: Pergamon, 1982.

\bibitem{Siegman}
A.\,E.~Siegman, Hermite-gaussian functions of complex argument as optical-beam eigenfunctions, J. Opt. Soc. Am. {\bf 63}, 1093 (1973).

\bibitem{Siegman97}
A.~Kostenbauder, Y.~Sun, and A.\,E.~Siegman, Eigenmode expansions using biorthogonal functions: complex-valued Hermite–Gaussians, J. Opt. Soc. Am. A {\bf 14}, 1780 (1997).

\bibitem{Lu}
B.~Lü, H.~Ma, A comparative study of elegant and standard Hermite–Gaussian beams, Opt. Commun. {\bf 174}, 99 (2000).

\bibitem{Allen}
L. Allen, M. W. Beijersbergen, R. J. C. Spreeuw, and J. P. Woerdman, Orbital angular momentum of light and the transformation
of Laguerre-Gaussian laser modes, Phys. Rev. A 45, 8185 (1992).

\bibitem{Case}
W.\,B.~Case, Wigner functions and Weyl transforms for pedestrians, Am. J. Phys. {\bf 76}, 937 (2008).

\bibitem{LL3}
L.D. Landau, E.M. Lifshitz, Quantum Mechanics (Pergamon Press, New York, 1991).

\bibitem{ST}
A.\,A.~Sokolov, I.\,M.~Ternov, Relativistic electron, Nauka, Moscow, (1974) [in Russian] [Radiation from Relativistic Electrons, Edited by C. W. Kilmister, American Institute of Physics Translation Series, New York, (1986)].

\bibitem{STJ}
A.\,A.~Sokolov, I.\,M.~Ternov, The Quantum Theory of the Radiating Electron, IV, Sov. Phys. JETP 1, 227 (1955).

\bibitem{Bagrov2002}
V.\,G.~Bagrov, M.C. Baldiotti, D.M. Gitman, and I.V. Shirokov, New solutions of relativistic wave equations in magnetic fields and longitudinal fields, J. Math. Phys. {\bf 43}, 2284 (2002).

\bibitem{Eseev}
M.\,K.~Eseev, I.\,N.~Meshkov, Traps for storing charged particles and antiparticles in high-precision experiments, Phys.-Usp. 59, 304 (2016).

\bibitem{McCulloch}
A.\,J.~McCulloch, D.\,V.~Sheludko, S.\,D.~Saliba, S.\,C.~Bell, M. Junker, K.A. Nugent and R.E. Scholten, Arbitrarily shaped high-coherence electron
bunches from cold atoms, Nat. Phys. 7, 785 (2011).

\bibitem{Jean}
O.~Jeandupeux, L.~Bürgi, A.~Hirstein, H.~Brune, K.~Kern, Thermal damping of quantum interference patterns of surface-state electrons, Phys. Rev. B 59, 15926 (1999).

\bibitem{GSI}
https://www.gsi.de/work/beschleunigerbetrieb/beschleuniger/ionenquellen/sources/ion\_\,sources.htm

\bibitem{Anderson}
S.\,E.~Anderson, G.~Raithel, Ionization of Rydberg atoms by standing-wave light fields, Nature Comm. 4, 2967 (2013).


\end{thebibliography}
\end{document}